\documentclass[aps,pre,twocolumn,showpacs]{revtex4-1}

\usepackage{amsfonts,amssymb,amsmath}

\usepackage{times}
\usepackage{graphicx}
\usepackage{bm}
\usepackage{color}

\newcommand{\bee}{\begin{equation}}
\newcommand{\eee}{\end{equation}}
\newcommand{\ba}{\begin{eqnarray}}
\newcommand{\ea}{\end{eqnarray}}

\newcommand{\ket}[1]{\vert #1 \rangle}
\newcommand{\bra}[1]{\langle #1 \vert}

\begin{document}
\title{\bf Length scales and self-organization in dense suspension flows}
\author{Gustavo D\"uring${}^{1,2}$, Edan Lerner${}^{1}$ and Matthieu Wyart${}^{1}$}

\affiliation {
${}^1$Center for Soft Matter Research, Department of Physics, New York University, New York, NY 10002, USA\\
${}^2$ Facultad de F\'isica, Pontificia Universidad Cat\'olica de Chile, Casilla 306, Santiago, Chile\\
}
\date{\today}

\begin{abstract}

Dense non-Brownian suspension flows of hard particles display mystifying properties: as the jamming threshold is approached, the viscosity diverges, as well as a length scale that can be identified from velocity correlations. To unravel the microscopic mechanism governing dissipation and its connection to the observed correlation length, we develop an analogy between suspension flows and the rigidity transition occurring when floppy networks are pulled -- a transition believed to be associated to the stress-stiffening of certain gels. After deriving the critical properties near the rigidity transition, we show numerically that suspension flows lie close to it. We find that this proximity causes a decoupling between viscosity and the correlation length of velocities $\xi$, which scales as the length $l_c$ characterizing the response to a local perturbation,  previously predicted to follow $l_c\sim 1/\sqrt{z_c-z}\sim p^{0.18}$ where $p$ is the dimensionless particle pressure, $z$ the coordination of the contact network made by the particles and $z_c$ is twice the spatial dimension. We confirm these predictions numerically, and predict the existence of a larger length scale $l_r\sim \sqrt{p}$ with mild effects on velocity correlation, and of a vanishing strain scale $\delta \gamma\sim 1/p$ that characterizes de-correlation in flow.
\end{abstract}


\maketitle

\section{Introduction}
Although the emergence of rigidity is generally associated with the breaking of a continuous symmetry, amorphous materials acquire rigidity without such a symmetry change, by jamming in a random  configuration. 
 Non-Brownian suspensions of hard particles are a particularly interesting example of this phenomenon, where the control parameter is the anisotropy of the applied stress \cite{gdrmidi}. As the stress anisotropy is reduced and the solid phase is approached, the dynamics is reminiscent of critical phenomena. Correlated motion, or eddies, appear in the flow, whose characteristic lengthscale $\xi$ seems to diverge \cite{pouliquen2004}. Concomitantly, rheological properties are singular \cite{gdrmidi,dacruz}, as characterized most accurately in suspensions \cite{boyer,lespiat} where the viscosity $\eta$ scales with the distance to jamming. Finally, a vanishing strain scale $\delta \gamma$ at which the structure reorganizes can be identified in experiments in which the direction of shear is reversed \cite{blanc}. Understanding these properties at a microscopic level remains a challenge. A visually attractive  image in  sheared granular materials is that the dynamics is governed by the buckling of forces chains, where stress is concentrated \cite{Tordesillas}. In another view dissipation is dominated by the formation of large eddies: assuming that the latter act as regions where particles move as blocs, particle velocities are amplified as $\xi$ increases, increasing dissipation and leading to a viscosity  $\eta\sim \xi^2$ \cite{heussinger2009}. 

To quantify and discuss the validity of these views, we consider the affine solvent model (ASM) \cite{durian95,olsson,hatano08a,heussinger2009,lernerpnas,andreotti} where hard particles cannot overlap, and receive a drag force proportional to their non-affine velocity, i.e.~their velocity with respect to an affinely-moving fluid background. The ASM captures the existence of a diverging length scale \cite{olsson,hatano08a,heussinger2009}, a vanishing strain scale \cite{olsson2010b}, and displays singular rheological properties \cite{olsson,lernerpnas} similar to experiments \cite{boyer}. In this model  an analogy was derived between the rheological properties of hard particles, and the elasticity of a  network made of identical springs, corresponding to the contacts between particles \cite{lernerpnas}. Using this analogy it was shown that if contact networks in flow were random, one would have $\eta\sim \xi^2\sim 1/(z_c-z)$ \cite{during12}, consistent with the idea that the length of eddies controls the viscosity. However, contact networks in flow are not random \cite{lernerpnas,Lerner2012}, but self-organize in configurations that are much harder to shear. The length scale $l_c$ characterizing the response of the flow velocity to a local perturbation was predicted to be insensitive to this self-organization and to follow $l_c\sim 1/\sqrt{z_c-z}$ \cite{during12},  while a simple assumption on the geometry of contact networks led to the prediction that  $\eta\sim(z_c-z)^{-2.7}$,  consistent with observations \cite{Lerner2012}.  It is presently unclear if the length scale $\xi$ describing velocity correlations is affected by this self-organization and is distinct from $l_c$, and if the viscosity is tied to $\xi$. The relationship between  self-organization and the  vanishing  strain scale at which configurations re-organize also remains not understood.

\begin{figure}[t]
\begin{center}
\centerline{\includegraphics[width=0.49\textwidth,trim=0mm 0mm 0mm 0mm, clip]{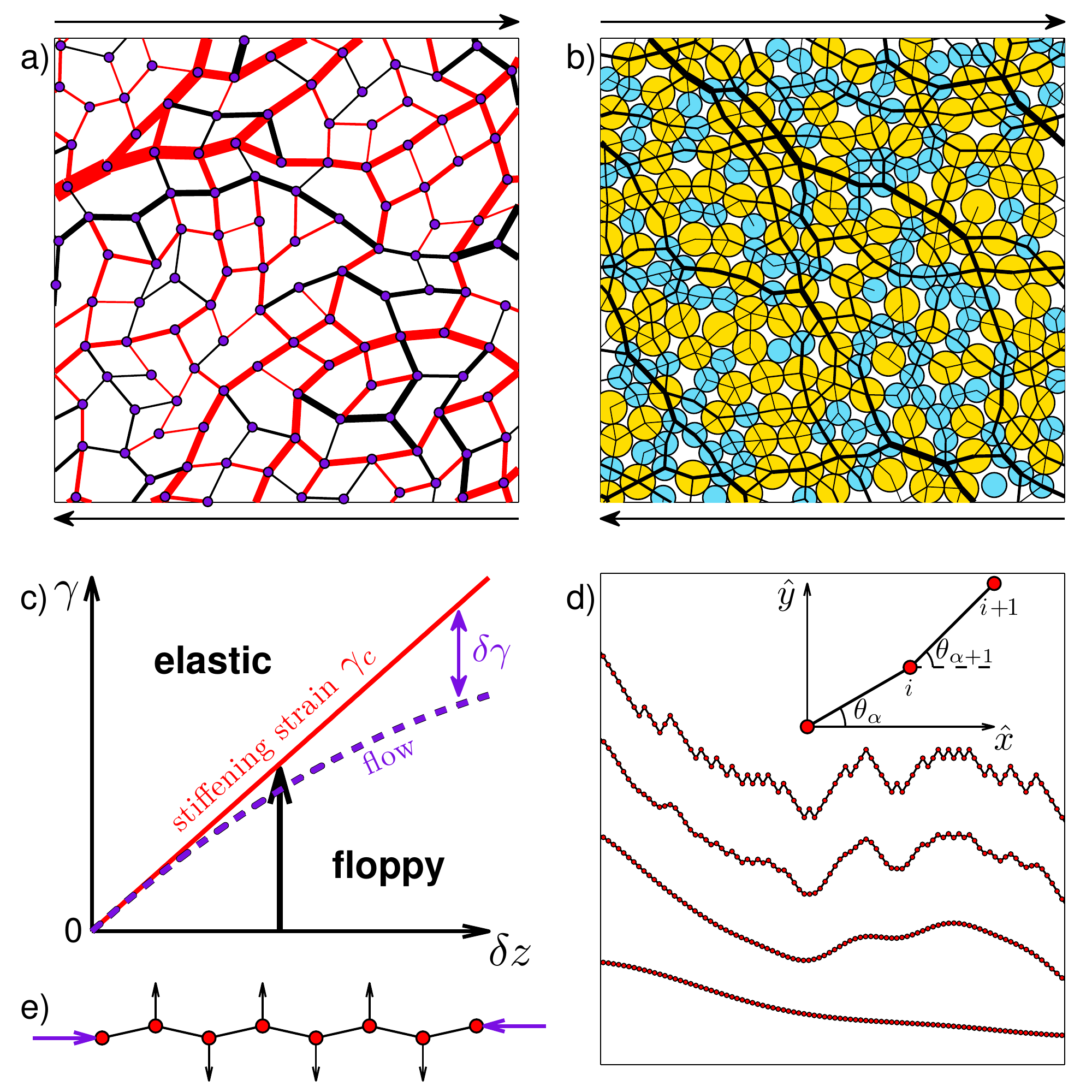} }
\caption{\label{net} (Color online) {\bf a)} A random floppy network under shear deformation approaching the SIJ transition; the thickness of the lines connecting the particles are proportional to the tension (red) or compression (black). Arrows indicate the direction of the shear.  {\bf b)} Snapshot of a contact force network in the ASM model of suspension flow, close but below the jamming threshold. {\bf c)}  Jamming phase diagram of networks in the coordination deficit $\delta z\equiv z_c-z$ and shear strain $\gamma$ plane, showing the line of the shear-induced transition (red solid line) $\gamma_c \sim \delta z$ \cite{Wyartmaha}.  Our contention  is that the contact networks in dense suspension  flows lie very close to this stiffening line, thus defining a characteristic strain   $\delta \gamma$  vanishing   at jamming. {\bf d)} A one dimensional chain under an extensional flow at four different times. The vertical position of the chain as been shifted for clarity, the upper curve represents the initial configuration. {\bf e)} An almost straight chain buckling under compressive flow.}
\end{center}
\end{figure}

 Here we shall argue that  flow self-organizes into configurations where
the buckling of force chains is very strong, leading to particle velocities and a divergence of the viscosity much larger than what naively 
expected from the size of the eddies $\xi$, which we find to scale as $l_c$. A simple illustration of the buckling mechanism is a nearly straight chain of rigid rods immersed in a viscous fluid, compressed at its two ends, as shown in Fig.~(\ref{net}.e). As we shall see, under certain circumstances when the line is almost straight the motion of the rods (and thus the  dissipation) can be huge, while  velocity correlations are short-ranged. We shall generalize the notion of buckling to floppy (under-coordinated)  networks of rigid rods where the dynamics drive the system away from a configuration in which contact forces are balanced. Generically, floppy networks with $z<z_c$ are unstable and cannot balance contact forces. However, as anticipated by Maxwell \cite{maxwell} and discussed by Calladine \cite{calladine}, certain floppy configurations can satisfy force balance. A simple way to generate such configurations is to pull on a floppy network until it jams, see for example a sheared network of rigid rods in Fig.(\ref{net},a), and the simpler case of a linear chain under extensional flow in Fig.(\ref{net},d). It has been proposed \cite{giessen,Wyartmaha,Mackintosh_PRE_2012,Frey_PRE_2007,Frey_PRL_2007} that this phenomenon generates the stress-stiffening of biopolymers networks \cite{louise} responsible for the non-linear response of certain biological tissues. This strain-induced jamming (SIJ) transition is the reverse process of what we will refer here to as the \emph{buckling transition}. 
 Apart from the case of certain crystals \cite{Sun}, these transitions are not well understood theoretically, even in simple models. In the first part of this manuscript we derive the scaling properties of the dynamics near these transitions, starting with the case of a line and extending our results to networks. A surprising result is that critical properties of  SIJ depend on initial conditions, whereas its reverse process, the buckling transition, is universal.  In a second part, we study the relationship between dense suspension flows and floppy networks pulled near jamming.   We show numerically that configurations in suspensions flows lie indeed very close to the buckling transition, as illustrated in  Fig.(\ref{net}c), and use this result to make several predictions for dense flows. 

\section{One dimensional chain} The case of a one dimensional chain already illustrates the difference between the dynamics under pulling toward jamming (the SIJ transition)  and pushing  away from it (buckling). It also captures several critical dynamical properties that also apply to flow. We first consider the SIJ transition of a linear chain of  $N$ massless nodes (particles) connected by rigid rods of size $a_0$, immersed in a two-dimensional  fluid of viscosity $\eta_0$, whose velocity profile corresponds to an extensional flow $\vec{\tilde{V}}_{\rm f}= \dot\gamma( x \hat{x} -y \hat{y})$, where $\dot{\gamma}$ is the strain rate. This flow induces a drag force on the particles $\vec{\tilde{F}}_{\rm drag}=\eta_0 (\vec{\tilde{V}}_i-\vec{\tilde{V}}_{\rm f})$, where $\vec{\tilde{V}}_i$ corresponds to the velocity of particle $i$. 
If $\tilde{\tau}_\alpha$ is the tension in the rod $\alpha$, balancing drag and contact forces reads:
\begin{equation}
\eta_0 (\vec{\tilde{V}}_i-\vec{\tilde{V}}_{\rm f})=\tilde{\tau}_{\alpha} {\vec n}_{ii-1} +\tilde{\tau}_{\alpha+1} {\vec n}_{ii+1}
\label{fb0}
\end{equation}
where ${\vec n}_{i j}$  is the unit vector going from node $i$ to $j$, and $\alpha$ links the nodes $i-1$ and $i$. 
To obtain dimensionless expressions we define the strain  $ \gamma \equiv t\dot{\gamma}$,  velocities $\vec{V}\equiv \vec{\tilde{V}}/(\dot{\gamma}a_0)$ and tensions $\tau \equiv \tilde{\tau}/(\dot{\gamma}a_0\eta_0)$. Introducing the angles $\theta_\alpha$ made by the rod $\alpha$ and the horizontal $x$ axis, as illustrated in the inset of Fig.(\ref{net}.d), we can rewrite Eq.(\ref{fb0}) by taking the difference of forces between two adjacent nodes, and projecting it along the unit vector ${\vec n}_{i j}$, and along the unit vector perpendicular to ${\vec n}_{i j}$,
\begin{eqnarray}
\label{forcebalance1}
&&f_{\rm ext} =\tau_{\alpha+1}\cos(\Delta \theta_{\alpha+1})-2\tau_\alpha+\tau_{\alpha-1}\cos(\Delta \theta_\alpha) \\
&&\dot\theta_\alpha +f^*_{\rm ext}= \tau_{\alpha+1}\sin(\Delta \theta_{\alpha+1})+\tau_{\alpha-1}\sin(\Delta \theta_\alpha).
\label{forcebalance2}
\end{eqnarray}
where $\Delta \theta_\alpha\equiv \theta_\alpha-\theta_{\alpha-1}$, $f_{\rm ext}\equiv \cos(2\theta_\alpha)$ and $f^*_{\rm ext}\equiv  \sin(2\theta_\alpha)$. Since the rods are rigid, the radial relative velocity $\dot{a}_0$ of the nodes vanishes in Eq.(\ref{forcebalance2}).  
The extensional flow stretches the chain until it undergoes a strain-induced jamming transition when $\theta_\alpha = 0$ for all $\alpha$, i.e.~it becomes a straight line. Appendix A shows that near the transition  the relative fluctuations of tension are negligible. Eq.(\ref{forcebalance1}) then implies:
\begin{equation}
\label{tau}
\tau \equiv \langle \tau \rangle = -\langle \Delta\theta^2\rangle^{-1}
\end{equation}
where $\langle\bullet \rangle\equiv\frac{1}{N}\sum_\alpha \bullet$. Using this result in Eq.(\ref{forcebalance2}) yields: 
\begin{equation}
\dot\theta_\alpha=(\theta_{\alpha+1}-2\theta_\alpha+\theta_{\alpha-1})/\langle \Delta\theta^2\rangle
\label{diffusionEquation}
\end{equation} 
Eq.(\ref{diffusionEquation}) is a  diffusion equation with a time-dependent diffusion constant. It implies that as the transition is approached, short wavelength fluctuations of the chain vanish first, as seen in Fig.(\ref{net}.d).
Thus a time-dependent length scale $l_{\rm corr}$ (measured in units of $a_0$) characterizes the shape, and most of the dynamics occur on that length scale.

\begin{figure}[t]
\begin{center}
\centerline{\includegraphics[width=0.49\textwidth]{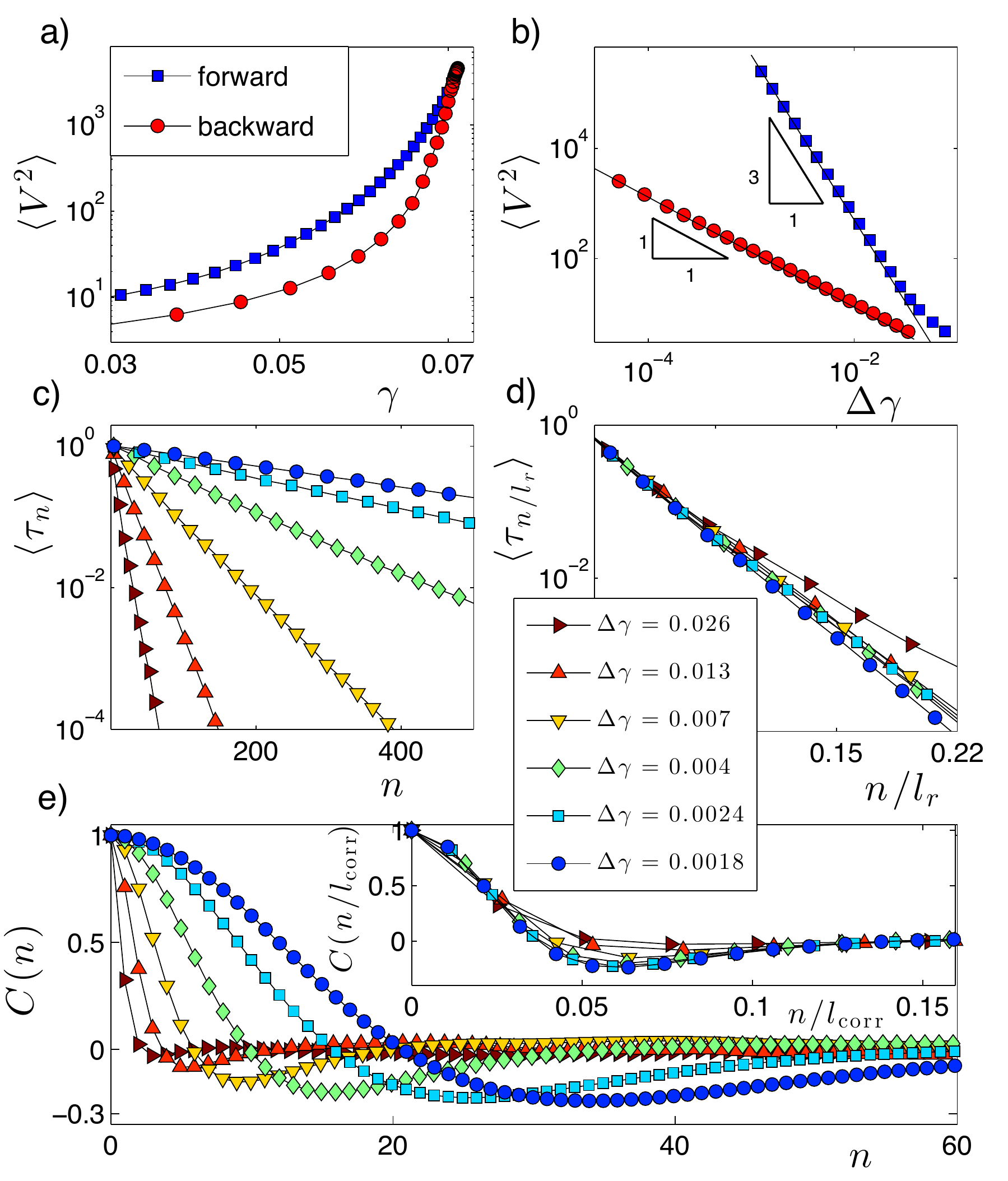}}
\caption{(Color online) Properties of a one-dimensional chain stretched by an extensional   flow with $N=10000$ under periodic boundary conditions, see Appendix~D for details on the numerical methods used. The initial distribution of the angles is random (i.e $\mu=0$).  {\bf a)}  $\langle V^2\rangle=\tau$ versus strain $\gamma$. The blue squares describe the evolution toward the strain-induced jamming
point and the red circles describe the buckling transition.  The reversibility of the backward evolution breaks at some strain $\gamma_{\rm noise}$. The only source of noise is round-off error. {\bf b)} Log-log plot of  $\langle V^2\rangle$  versus the distance to the critical point  $\Delta \gamma$. The predicted exponent are  $3$ and $1$ in perfect agreement with the simulations {\bf c)} Average response to a local perturbation $\langle \tau_n\rangle$ {\it v.s.} the the distance $n$ to the perturbation. {\bf d)} Rescaled average response  $\langle \tau_n\rangle$ {\it v.s.} rescaled distance $n/l_r\sim n \Delta \gamma^{3/2}$. {\bf e)} Average velocity correlation $C(n)$ vs the distance $n$. { Inset:} Rescaled  average velocity correlation $C(n)$ vs rescaled distance  $n/l_{\rm corr}\sim n \Delta \gamma$. \label{linePro}}
\end{center}
\end{figure}

Eq.(\ref{diffusionEquation}) can be solved by introducing the Fourier decomposition $\theta_\alpha(t)=\frac{1}{N}\sum_{k=0}^{N-1}\theta_k(t) e^{\frac{2\pi i}{N}k\alpha}$, leading to a closed equation for $\langle \Delta\theta^2\rangle$:
\begin{equation}
\langle \Delta\theta^2\rangle=\frac{4}{N^2}\sum_{k} \vert \theta_{k}(0)\vert^2e^{-2\nu_k\int_0^{\gamma}  \frac{d\tau}{\langle \Delta\theta^2\rangle}}\sin[\pi  k/N]^2
\label{self}
\end{equation}
where  $\nu_k=2(1-\cos[\frac{2\pi k}{N}])$.  Eq.(\ref{self}) depends on the initial angles $\theta_k(0)$ . To illustrate this dependence,  we  consider random initial conditions whose Fourier component are independent and present a well-defined second moment:   $\overline{ \vert \theta_{k}(0)}\vert^2/N=g(k)$, where $\overline{\bullet}$ denotes averaging over the initial conditions distribution in the thermodynamic limit. We consider  the family of initial conditions defined as   $g(k)=\epsilon^2 c_\mu \sin(\pi k)^{\mu}$, with $\mu$ a non-negative even exponent and $c_\mu$ a normalization constant such that $\int_0^1g(k)dk=\epsilon^2$, where $\epsilon$ is the average amplitude of the initial angles.  $\mu=0$ implies $g(k)=\epsilon^2$, meaning that the chain is initially a random walk.  For  $\mu\rightarrow\infty$, one gets $g(k)=\epsilon^2 \delta(k-1/2)$ which corresponds to a crystal with long range order. For given initial conditions, one can solve the asymptotic behavior  of Eq.(\ref{self}) in the thermodynamic limit, to obtain:
\begin{eqnarray}
\langle \Delta\theta^2\rangle&\propto& (\gamma_c-\gamma)^{\frac{3+\mu}{1+\mu}} \label{omegamin},\\
\langle V^2\rangle= \tau &\propto& (\gamma_c-\gamma)^{-\frac{3+\mu}{1+\mu}}, \label{V}\\
\langle \dot\theta^2 \rangle&\propto&  (\gamma_c-\gamma)^{-1},\label{V2}
\end{eqnarray}
where  $\gamma_c=\epsilon^2/2$. The equality in Eq.(\ref{V}) stems from balancing the invested work per unit time per unit length $\eta_0  \sum_i \tilde{V}_i^2/ a_0N= \eta_0\dot{\gamma}^2a_0 \langle V^2\rangle$  with the dissipation rate per unit length $\tilde{\tau}\dot\gamma= \eta_0\dot{\gamma}^2a_0 \tau$.  The scaling of these quantities results from Eqs.(\ref{tau},\ref{omegamin}).
The third equation can be obtained from the simple geometrical consideration that the additional horizontal strain needed to reach the transition is  $\gamma_c-\gamma=1-\langle \cos(\theta_\alpha)\rangle$.
The results are that (i) initial conditions affect the critical properties and (ii) relative velocities, which are of the order of~$\dot\theta$ (Eq.(\ref{V2})), and absolute velocities $V$ (Eq.(\ref{V})), do not scale in the same way in general, except for $\mu=\infty$. The prediction for the scaling of $\langle V^2\rangle$ is tested in Fig.(\ref{linePro}.b).

The diffusive dynamics implies that a  length scale $l_{\rm corr}$ grows with strain. This length  is picked up by velocity correlations, which can be obtained from Eqs.(\ref{diffusionEquation},\ref{omegamin}):
\begin{eqnarray}
C(n)&\equiv&\langle {\vec V}_{i}\cdot {\vec V}_{i+n}\rangle\sim \left(\frac{n}{l_{\rm corr}}\right)^{2+\mu}e^{- (n/l_{\rm corr})^2},\\
l_{\rm corr}&\sim& \left(\int_0^t  \frac{d\tau}{\langle \Delta\theta^2\rangle}\right)^{1/2}\sim(\gamma_c-\gamma)^{-\frac{1}{1+\mu}},
\end{eqnarray}
in good agreement with the results of Fig.(\ref{linePro},e).

A second and larger length scale diverges near the transition, and can be identified from the response to a local perturbation, defined as follows:  at some time the driving flow is stopped, and one single contact $\beta$ is elongated at some rate, leading to a dipole of force of amplitude $\tau_0$ on the two nodes adjacent to $\beta$.  The  response to such a local perturbation is described by Eq.(\ref{forcebalance1}), where the external force $f_{\rm ext}$ is now $f_\alpha=\delta_{\alpha,\beta} \tau_0$.  Eq.(\ref{forcebalance1}) can be written in  matrix form using the braket notation
 $\tau_0 \ket{\beta}=-\mathcal{N}\ket{\tau}$
where $\bra{\alpha}\tau\rangle=\tau_\alpha$, $\bra{\alpha}\beta\rangle=\delta_{\alpha,\beta}$ and the $\mathcal{N}$-matrix is given by
\begin{equation}
\mathcal{N}_{\alpha \beta}=2\delta_{\alpha, \beta}-\cos(\theta_\alpha-\theta_\beta)(\delta_{\beta, \alpha+1} +\delta_{\beta, \alpha-1} ).
\label{N1D}
\end{equation}
Inverting the $\mathcal{N}$-matrix is difficult  in general. However, close to the critical point Appendix A shows that  $\mathcal{N}$ can be approximated by its average value $\mathcal{N}_a\equiv\langle\mathcal{N}\rangle$. Translational invariance implies that the eigenvectors of $\mathcal{N}_a$ are plane wave modes, with eigenvalues $\omega^2_q=2(1-(1-\langle \Delta\theta^2\rangle/2)\cos(2\pi q/N))$ where $q$ is the wavenumber. The point response then reads 
$$ \tau_{n} = -\tau_0\bra{\beta + n}\mathcal{N}_a^{-1}\ket{\beta} \sim\tau_0 \sum_{q} e^{i2\pi q n}/\omega^2_q\sim \tau_0 e^ {-\vert n\vert / l_{r}}$$
where $n$ is the distance to the perturbation, and:
\begin{equation}
\label{13}
l_{r}=1/\sqrt{\langle \Delta\theta^2\rangle}\sim (\gamma_c-\gamma)^{-\frac{3+\mu}{2+2\mu}}
\end{equation}
is measured in units of $a_0$. Our third key result is thus that the SIJ transition in the linear chain is characterized by two length scales: $l_{\rm corr}$ which characterizes the velocity correlations, and $l_r$ which characterizes the response to a local perturbation. These lengthscales are tested numerically in Fig.(\ref{linePro},c,d,e).

{\it Buckling}: The dynamics described by Eq.(\ref{diffusionEquation}) is formally reversible. However, we find that if the strain rate is reversed at some  strain $\gamma_f$  smaller than but close to $\gamma_c$, the observed trajectory is not reversible, as shown in Fig.(\ref{linePro}a,b). Instead, in this backward evolution the dynamics becomes dominated by high spatial frequencies of the structure, and is  well described by Eqs.(\ref{omegamin},\ref{V},\ref{V2},\ref{13}) with $\mu=\infty$ and some effective critical strain $\tilde \gamma_c<\gamma_c$, as shown in Appendix B. The correlation length follows $ l_{\rm corr}\approx 1$. All these results are independent of the initial conditions: in contrast with the forward process, the reverse process is universal.

These results stem from the simple fact that when time is reversed, the tension in the rods changes sign (i.e.~becomes compressive) and the diffusion coefficient of Eq.(\ref{diffusionEquation}) becomes negative, which results in the amplification of high frequency noise. Physically, this means that upon reversing the strain rate, the chain buckles on the smallest available length scale, leading to the universal results $\langle \Delta\theta^2\rangle= \tilde\gamma_c-\gamma$, where  $\tilde\gamma_c-\gamma_f\propto \langle \Delta\theta(\gamma_{f})^2\rangle$, as shown in Appendix B.

\section{Networks}

\begin{figure}[t]
\begin{center}
\centerline{\includegraphics[width=0.5\textwidth,height=0.195\textwidth]{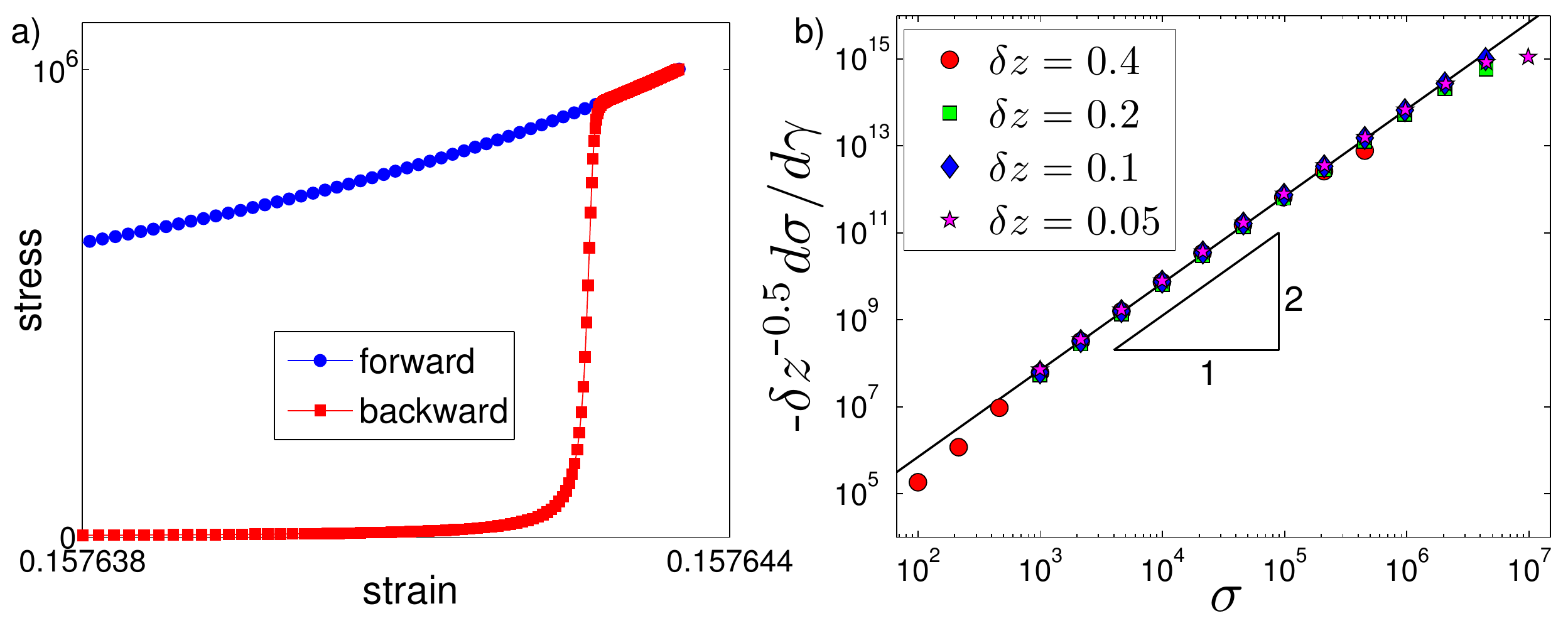} }
\caption{(Color online) 2-D network, built as in \cite{Wyartmaha}, under shear. {\bf a)} stress $\sigma$ vs strain $\gamma$. The blue circles represent the dynamics towards the SIJ transition, whereas the red squares represent the buckling transition. {\bf b)} Scaling of $\frac{d\sigma}{d \gamma}/\delta z^{0.5}$ vs $\sigma$ in the vicinity of the buckling transition (see Appendix~D for details about numerical methods).
 The slope of the line is the predicted exponent $2$. \label{reverseStrain} }
\end{center}
\end{figure}

Our results on the linear chain apply more generally to networks of $N_c$ rigid rods connecting $N$ nodes in spatial dimension $d$, immersed in a flowing solvent; see an example in Fig.(\ref{net},a). Unlike the case of the linear chain, in networks the coordination $z=2N_c/N$ can be manipulated.  Varying this parameter allows us to study the interplay between the rigidity transition that occurs in disordered isotropic networks when springs are added up to $z_c$, which is known to be associated with a diverging length scale $l_c\sim 1/\sqrt{\delta z}$ \cite{during12}, and the SIJ transition. We consider floppy networks having $\delta z \equiv z_c - z >0$; it has been shown that when deformation is imposed on such networks, they eventually undergo a SIJ transition at some strain $\gamma_c \sim \delta z$ \cite{Wyartmaha}. At $\gamma_c$ the velocities diverge, which implies a divergence of the stress and  tension, as in the linear chain case. We focus on the buckling process. As in the case of the linear chain (compare Fig.(\ref{reverseStrain},a) with Fig.(\ref{linePro},a)), a time-reversed diffusive process is the source of the irreversibility upon the backward dynamics. We thus expect this process to follow the same dependence on strain as the buckling of a linear chain,  in particular that the dimensionless shear stress $\sigma\equiv \tilde{\sigma}/(a_0\eta_0\dot\gamma)$ (which plays the role of the dimensionless tension $\tau$ for the chain) follows $\sigma \sim C(\delta z)/(\gamma_c - \gamma)$, or equivalently $d\sigma/d\gamma \sim -\sigma^2/C(\delta z)$, where $C(\delta z)$ is some function whose derivation is left for future work. These predictions are tested numerically in random networks of rigid rods under simple shear with the driving flow $\vec{\tilde{V}}_{\rm f}=\dot\gamma y \hat{x}$ (see Appendix~D for numerical methods). Results are shown in Fig.(\ref{reverseStrain},b), which supports that $C(\delta z)\sim 1/\delta z^{0.5}$.

\subsection{Spectrum} 
\begin{figure}
\begin{center}
\centerline{\includegraphics[width=0.45\textwidth]{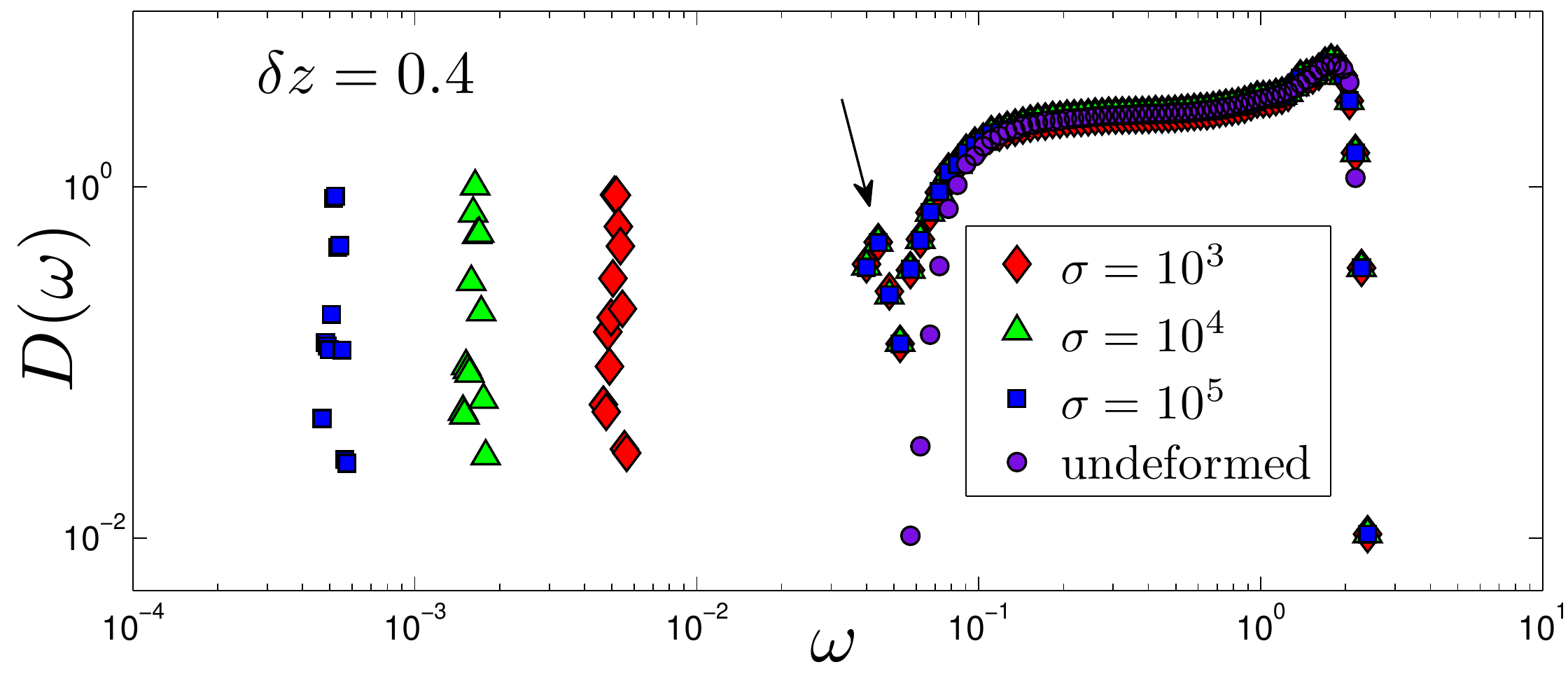}}
\caption{ (Color online) The spectral density $D(\omega)$ of the ${\cal N}$ operator for 2d networks of size $N=4096$, measured numerically for $\delta z = 0.4$ at various stresses, as indicated by the legend. In undeformed networks a gap is observed in $D(\omega)$ up to a frequency $\omega^* \sim \delta z$ \cite{during12}. Above $\omega^*$, $D(\omega)$ remains effectively unchanged under the deformation, whereas below $\omega^*$ a single mode of frequency $\omega_{\rm min}$  rises from zero as $1/\sqrt{\sigma}$ (at constant z) upon buckling, for which $\sigma \to \infty$. Notice that $D(\omega)$ displays modes between $\omega_{\rm min}$ and $\omega^*$ indicated by the arrow; these are plane-wave modulations of the minimal mode, and thus scale like $1/L$.
\label{Density}}
\end{center}
\end{figure}

In the buckling process, we expect that the structural length scale $l_{\rm corr}$ plays no role, as for the linear chain. We are thus left with two length scales at play: the length $l_r$ associated with the SIJ transition, and $l_c$ associated with the rigidity transition.
To disentangle the role of these,  it is convenient to extend the definition of the linear operator ${\cal N}$  introduced for the chain in Eq.(\ref{N1D}) to higher dimensional networks.  Following Calladine \cite{calladine}, we first introduce  the operator $\mathcal S$ of dimension $N_c\times Nd$, with components $\mathcal{S}_{\alpha k}=(\delta_{j,k}-\delta_{i,k}){\vec n}_{ij}$, where $\alpha$ labels the rod connecting the node $i$ with node $j$. It is easy to check that $\mathcal S^T|\tilde{\tau}\rangle\equiv |\tilde{F}_u\rangle$  is the set of unbalanced forces appearing on the nodes if the rod contact forces are $\ket{\tilde{\tau}}$, whereas $S|\tilde{V}\rangle$ is the rate at which rods change length if nodes were moving  with a velocity $|\tilde{V}\rangle$ \cite{calladine}. In our notations $\vec{\tilde{V}}_i=\bra{i}\tilde{V}\rangle$. If a fluid with a velocity field $\ket{\tilde{V}_{\rm f}}$ is creating a drag force $|\tilde{F}_{\rm drag}\rangle = \eta_0(\ket{\tilde{V}_{\rm f}}-\ket{\tilde{V}})$ on the nodes,
it is easy to show, see  \cite{lernerpnas} or Appendix C, that the contact forces $\ket{\tilde{\tau}}$ appearing on the rods follow:
\begin{equation}
-\eta_0\mathcal S\ket{\tilde{V}_{\rm f}}=\mathcal{N}\ket{\tilde{\tau}}.
\label{forcehigh1}
\end{equation}
where  $\mathcal{N}\equiv\mathcal{S}\mathcal{S}^T$.
$\mathcal{N}$ is symmetric and can be diagonalized $\mathcal{N}=\sum_\omega \omega^2 |\phi(\omega)\rangle\langle\phi(\omega)|$.  Eq.(\ref{forcehigh1}) shows that  the contact forces, and hence the stress, can diverge only if there are modes approaching zero-frequency in ${\cal N}$. The spectral density $D(\omega)$ in isotropic random floppy networks can be computed theoretically; it presents a gap at some frequency $\omega^*\sim \delta z$, above which it plateaus as shown in Fig.(\ref{Density}), and the characteristic scale of the modes at $\omega^*$ is $l_c\sim 1/\sqrt{\delta z}$ \cite{during12}.

%

As the anisotropy increases additional features must enter in the spectrum. At $\gamma_c$ a set of self-balancing rod forces $\ket{\phi_0}$  appears, which constitutes a zero-mode of $\mathcal{N}$. Additional modes denoted $| \phi_{\omega(k)}\rangle$ must be present below $\omega^*$, which correspond approximatively to plane-wave modulations of $\ket{\phi_0}$ \cite{Lerner2012}: $\langle\alpha\ket{\phi_{\omega(k)}}\approx \langle\alpha\ket{\phi_0}\exp(i 2\pi \vec{r}_{\alpha}\cdot\vec{k})$, where $\vec{r}_{\alpha}$ is the position of contact $\alpha$. Their frequency follows $\omega(k)^2=\bra{\phi_{\omega(k)}}\mathcal{N}\ket{\phi_{\omega(k)}}\approx A k^2$ where $A$ is a constant, leading to a  Debye density of states below $\omega^*$ with $D(\omega)\sim \omega^{d-1}$. In our finite-size simulations only a few such modes can be visualized; an example is indicated by an arrow in Fig.(\ref{Density},a). 
 
 When buckling occurs as $\gamma$ decreases below $\gamma_c$, contact forces are not balanced anymore, and the minimal eigenvalue $\omega_{\rm min}^2$  of ${\cal N}$ rises from zero as shown in Fig.(\ref{Density}). Near $\gamma_c$ the stress will be governed by this minimal mode $|\phi_0\rangle$, as can be deduced from a spectral decomposition of Eq.(\ref{forcehigh1}), implying that the norm of the dimensionless contact forces $\tau\equiv \tilde{\tau}/(a_0\eta_0\dot\gamma)$ follows $||\tau||\sim1/\omega_{\rm min}^2$ (with some coordination-dependent prefactor). Since shear-stress and contact forces are proportional, we obtain $\omega_{\rm min}\sim C_2(\delta z)/\sqrt{\sigma}$ where $C_2(\delta z)$ is some function. We find numerically and will justify elsewhere  that for isotropic networks that are pulled near the SIJ transition, $C_2(\delta z)\sim \sqrt{\delta z}$. As for the zero mode at $\gamma_c$, there are also modes in the spectrum of ${\cal N}$ corresponding to plane-wave modulation of the minimal mode, easily shown to lead to a density of states $D(\omega)\sim \omega(\omega^2-\omega_{\rm min}^2)^{(d-2)/2}$ for $\omega<\omega^*$. 

\subsection{Length scales} We can now consider the point response of networks, obtained by changing the length of a single rod $\beta$ at a rate of $t_0^{-1}$ at a fixed strain $\gamma$. This procedure corresponds to replacing  $S\ket{\tilde{V}_{\rm f}}$ by $t_0^{-1}|\beta\rangle$ in Eq.(\ref{forcehigh1}), giving rise to the rod forces $|\tilde{\tau}\rangle=\eta_0t_0^{-1} \mathcal{N}^{-1} |\beta\rangle$. Consequently, the dimensionless force $\tau_\alpha \equiv \tilde{\tau}_\alpha t_0/\eta_0$ in a rod $\alpha$ due to elongating the rod $\beta$ is given by $\langle \alpha \ket{\tau} = \bra{\alpha}{\cal N}^{-1}\ket{\beta}$. Using an eigenvalue decomposition of the ${\cal N}$ operator, we identify and collect separately the contributions to $\langle \alpha \ket{\tau}$ stemming from modes having frequencies $\omega>\omega^*$ and $\omega<\omega^*$, denoted in the following as $\langle \alpha \ket{\tau}_*$ and $\langle \alpha \ket{\tau}_{\rm min}$. The modes above $\omega^*$ are statistically equivalent to those of  undeformed random networks, for which it is known that $\langle \alpha \ket{\tau}_* \sim e^{-r/ l_c}$  \cite{during12} with an additional algebraic pre-factor of $r$ that depends on the spatial dimension,  where $r \equiv \vert \vec{r}_{\alpha}-\vec{r}_{\beta}\vert$ and $l_c \sim \delta z^{-1/2}$.

The contribution stemming from the modes below $\omega^*$ can be calculated using the modes $\ket{\phi_{\omega(k)}}$ introduced above, leading to
$$ \langle\alpha\ket{\tau}_{\rm min}=\sum_{\omega(k)<\omega*} \frac{\langle \phi_0\ket{\beta}\langle\alpha\ket{\phi_0}\exp(i 2\pi (\vec{r}_{\alpha}-\vec{r}_{\beta})\cdot\vec{k})}{\omega_{\rm min}^2+A\,k^2}.$$
The sum over these plane-wave modes is dominated by the low wave-numbers $k$. Taking the continuum limit, we obtain an exponential decay in the response for large $r$:
\begin{equation}
\langle\alpha\ket{\tau}_{\rm min}\sim\langle \phi_0\ket{\beta}\langle\alpha\ket{\phi_0} e^{-r/l_r},
\end{equation}
where $l_r\sim1/\omega_{\rm min}$. The factor  $\langle\alpha\ket{\phi_0}$  can fluctuate from rod to rod but since $\ket{\phi_0}$ is an extended mode it does not decay with the distance $r$.

Both contributions $\langle\alpha\ket{\tau}_{\rm min}$ and  $\langle\alpha\ket{\tau}_*$ have exponential decays, but on different length scales $l_r$ and $l_c$. To determine which is the dominant length scale, we need  to  estimate the relative amplitude of each contribution.  This can be done by calculating the total norm  of the rod forces $\vert\vert \tau \vert\vert^2$ which reads
\begin{equation}
\bra{\tau}\tau\rangle=\bra{\beta}\mathcal{N}^{-2}\ket{\beta}=\!\underbrace{\sum_{\omega^*>\omega} \!\!\frac{\langle \phi(\omega)|\beta\rangle^2}{\omega^4} }_{\tau^2_{\rm min}}+\!\!\underbrace{\sum_{\omega^*<\omega} \!\!\frac{\langle \phi(\omega)|\beta\rangle^2}{\omega^4}}_{\tau^2_*}. \label{forceResponse}
\end{equation}
Using  $\sum_{\beta}{\langle \phi(\omega)|\beta\rangle^2}=1$, one can express the average of these norms in the thermodynamic limit in terms of $D(\omega)$, 
leading to $\langle\tau^2_*\rangle\sim 1/\delta z^3$ and $\langle\tau^2_{\rm min}\rangle\sim\omega^{d-4}_{\rm min}$. 
We conclude that if $\delta z<\omega_{\rm min}^{(4-d)/3}$, the response to a local perturbation has an initial exponential decay on the scale $l_c$, which then crosses over to a much slower decay on the scale $l_r$ for larger distances. However close enough to the buckling transition, i.e. if  $\omega_{\rm min}^{(4-d)/3}< \delta z$, the response in force is entirely controlled by $l_r$.  

\begin{figure}[t]
\begin{center}
\centerline{\includegraphics[width=0.46\textwidth]{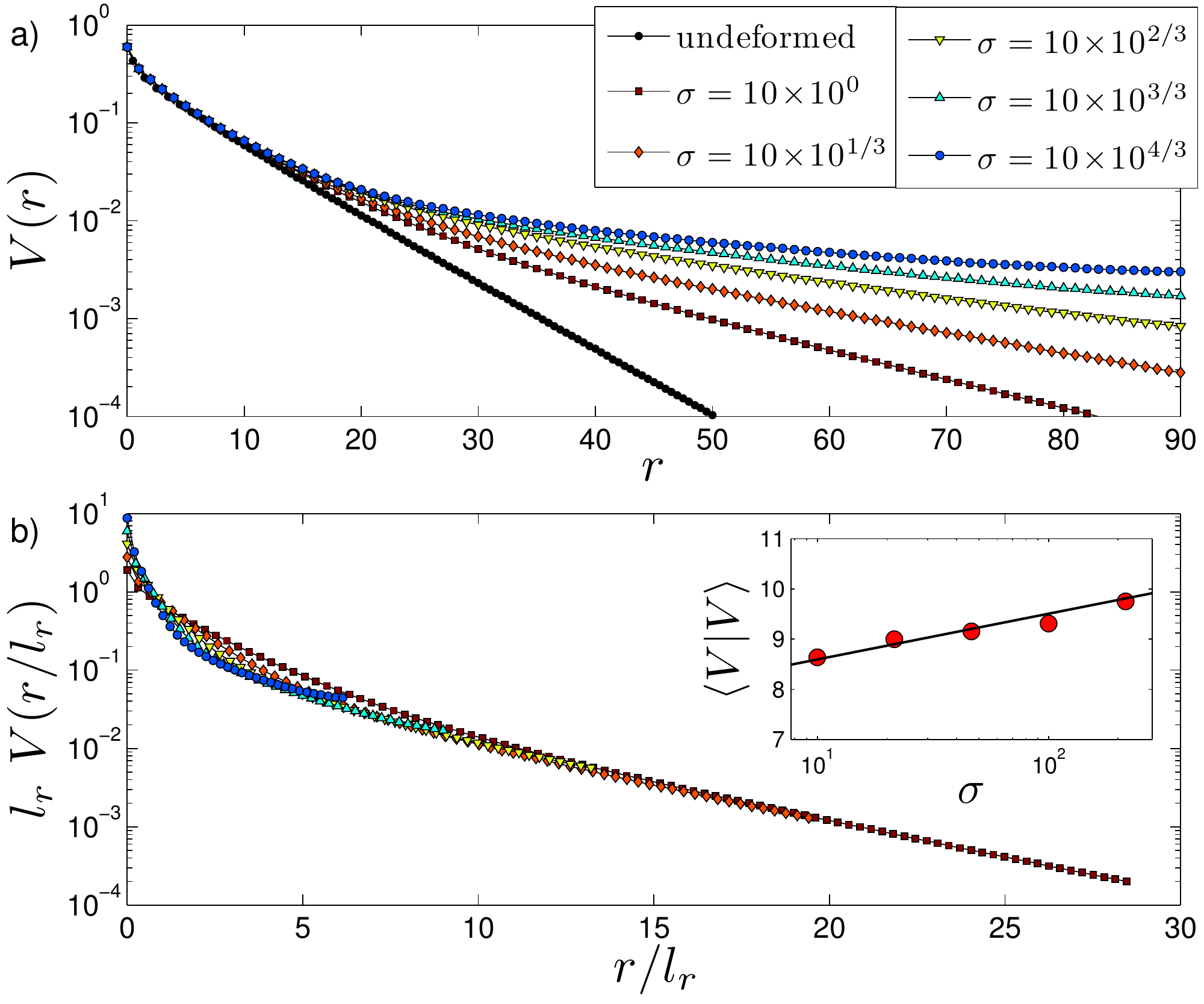}}
\caption{ (Color online) a) Velocity field amplitude $V(r)$ as a response to a local perturbation \emph{vs.}~the distance to the perturbation $r$ for different shear stress in two dimensions. 
b) Rescaled velocity $V(r) \sqrt\sigma$  {\it vs} rescaled distance $r/\sqrt \sigma$. The collapse implies that at fixed coordination $l_r\sim\sqrt{\sigma}$ as predicted.  Inset: the total velocity field displacement $\langle V\vert V\rangle$ \emph{vs}.~the stress $\sigma$, showing the predicted logarithmic increase.\label{VelResponse}}
\end{center}
\end{figure}

The dimensionless velocity field $\ket{V} = \frac{t_0}{a_0}\ket{\tilde{V}}$ in response to the same perturbation can also be calculated. This response must also have  two terms decaying exponentially with length scales $l_r$ and $l_c$, but the relative amplitudes of these terms turns out to differ from that of the force response. To determine the amplitude of each contribution, we estimate, as for the force response Eq.(\ref{forceResponse}), their respective roles in the total norm of the velocity  field $ \bra{V}V\rangle=\bra{\beta}\mathcal{N}^{-1}\ket{\beta}= V^2_{\rm min}+ V^2_*$. We find $\langle V^2_*\rangle\sim 1/\delta z$ and $\langle V^2_{\rm min}\rangle \sim \log(\omega^*/\omega_{\rm min})$ in $d=2$, or $\langle V^2_{\rm min}\rangle \sim \delta z^{d-2}$ in larger dimension. Thus in terms of velocities for $d>2$, the dominant response always decays as $l_c$, and $l_r$ can only be observed far from the deformed contact. Both the length scale and the amplitude of the response are in excellent agreement with the numerics (see Fig.(\ref{VelResponse})) that reflect both the presence of strain-independent length $l_c$, as well as a diverging length $l_r$. 

Velocity correlations near the buckling transition are associated with the properties of the minimal mode $|\phi_0\rangle$, 
which are analytically hard to access. However, large networks can be simulated allowing to measure correlations adequately. We find that the correlations of velocities under shear are very similar to the velocity response to a point perturbation, as shown in Appendix E: they present an initial decay on the length $l_c$, followed by a decay of much milder amplitude on the length $l_r$, that is visible only for $r>l_c$.  Thus networks near buckling present a clear decoupling between the dimensionless stress -diverging near buckling- and velocity correlations,  mainly governed by $l_c$ which is independent from the distance to the buckling transition.

\begin{figure}[t]
\begin{center}
\centerline{\includegraphics[width=0.5\textwidth]{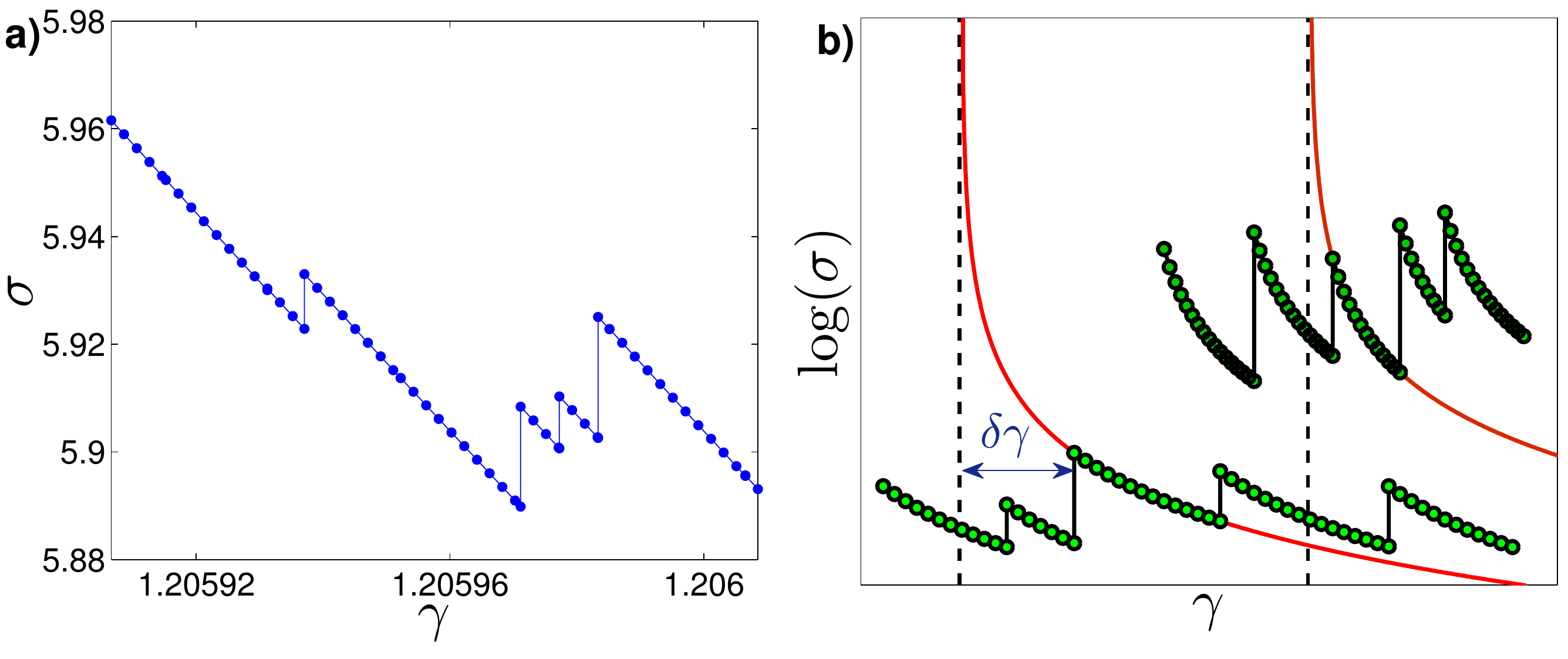}}
\caption{ (Color online) Typical stress \emph{vs}.~strain signal from the ASM simulations \cite{lernerpnas,simulation_paper}. The stress relaxes between the abrupt increases which signal collisions of the sheared hard particles. The slope of the stress during the relaxation segments depends on the stress itself as $d\sigma/d\gamma \sim -\sigma^2$, as derived in Appendix~C, and validated numerically in Fig.(\ref{flowProp},b). {\bf b)} Illustration of the relation between the stress relaxation in between collisions in the ASM, and the buckling transition: the stress follows $\sigma \sim (\gamma-\gamma_c)^{-1}$ (red continuous curves) between collisions, which is analogous to the stress relaxation of floppy networks evolving \emph{away} from a buckling transition. The dashed vertical lines indicate the strain $\gamma_c$ at which a floppy network -- constructed by connecting the centers of the hard particles in contact by rods -- would undergo a SIJ transition if it were sheared backwards. We note that while the strain intervals between collisions (the length of the dotted curves) vanish in the thermodynamic limit, the strain scale $\delta\gamma \equiv \gamma-\gamma_c \sim 1/\sigma$ is independent of system size. \label{stress_relaxation_fig}}
\end{center}
\end{figure}

\section{Suspension flows}
 
In the ASM of suspension flows \cite{durian95,olsson,hatano08a,heussinger2009,lernerpnas,andreotti} contacts made between particles define a network, as shown in Fig.(\ref{net},b). 
 The topology of the network evolves as contacts break and are formed during collisions \cite{lernerpnas}.  At fixed volume and constant $\dot \gamma$, at discrete instants of time, contacts are formed, leading to abrupt jumps in the stress (infinitesimal as $N\rightarrow \infty$), which relaxes between collisions, as shown in Fig.(\ref{stress_relaxation_fig},a). We make the hypothesis that the smooth motion in between collisions is dominated by the proximity of a buckling transition, leading to a decoupling between viscosity and the velocity correlation length. 
According to this hypothesis, our results on constructed networks near their buckling transition must apply to flow: no structural length $l_{\rm corr}$ should play a role, but the two lengths $l_c$ and $l_r\sim 1/\omega_{\rm min}$ should enter in velocity correlations, whereas $\sigma\sim C'(\delta z)/(\gamma-\gamma_c)$ and $\omega_{\rm min}\sim C_2'(\delta z)/\sqrt{\sigma}$. The functions $C'(\delta z)$ and  $C_2'(\delta z)$ can in principle differ from $C(\delta z)$ and  $C_2(\delta z)$ introduced for networks because aspects of the respective networks are different: in flow (i)  contact forces must be positive, and (ii) near jamming the contact network has a non-vanishing anisotropy, i.e.~the shear component of the stress does not vanish \cite{lernerpnas}. Both facts are in contrast to our constructed networks, for which forces can be negative, and for which the  anisotropy vanishes even at $\gamma_c$ as $\delta z\rightarrow 0$ \cite{Wyartmaha}. From these considerations, together with the additional hypothesis that  non-affine displacements and relative displacements between particles in contact are of the same order, it is straightforward to show (see Appendix C) that $C'(\delta z)\sim \delta z^0$ and $C_2'(\delta z) \sim \delta z^0$. These results imply that $d\sigma/d\gamma\sim -\sigma^2$, as confirmed numerically in Fig.(\ref{flowProp},b) (see Appendix~D for details about numerical methods). These predictions can also be tested by  replacing contacts between particles in flow at some instant of time by rigid rods,  and by shearing the resulting network {\it backward}, until  a SIJ transition  occurs. We predict a transition for a backward strain of amplitude $\delta \gamma\equiv (\gamma-\gamma_c)\sim 1/\sigma$, as illustrated in Fig.(\ref{stress_relaxation_fig},b). This result is checked in Fig.(\ref{flowProp},a). Since observations in flow indicate $\sigma \sim\delta z^{-2.7}$ \cite{lernerpnas}, we get $\delta \gamma\sim \delta z^{2.7}\ll\delta z$.  These results support  the hypothesis that suspension flows lie much closer to a buckling transition than isotropic random networks for which $\delta \gamma\sim\delta z$. It also yields a characteristic strain scale $\delta \gamma\sim 1/\sigma$, in agreement with the numerical observation that particle velocities decorrelate on that strain scale \cite{olsson2010b}.

Our scenario imposes that the spectrum of $\mathcal N$ presents a minimal mode at a frequency $\omega_{\rm min}\sim 1/\sqrt{\sigma}$, as previously observed \cite{lernerpnas} and argued for \cite{Lerner2012}. Thus, following our results of the previous section on networks, we expect both the response to a local perturbation and the velocity correlations to be  characterized by two length scales: $l_c\sim1/\sqrt{\delta z}$, and a much larger $l_r\sim\sqrt{\sigma}$ that should govern correlations only for $r\gg l_c$. Fig.(\ref{flowProp}) confirms our central result that both the response to a point perturbation and velocity correlations decay as $e^{-r/l_c}$, as expected for $r \lesssim l_c$. For $r\gg l_c$ however, the characteristic decay length should be $l_r$, but this prediction is hard to test numerically as our system sizes are limited. 

\begin{figure}[t]
\begin{center}
\centerline{\includegraphics[width=0.45\textwidth]{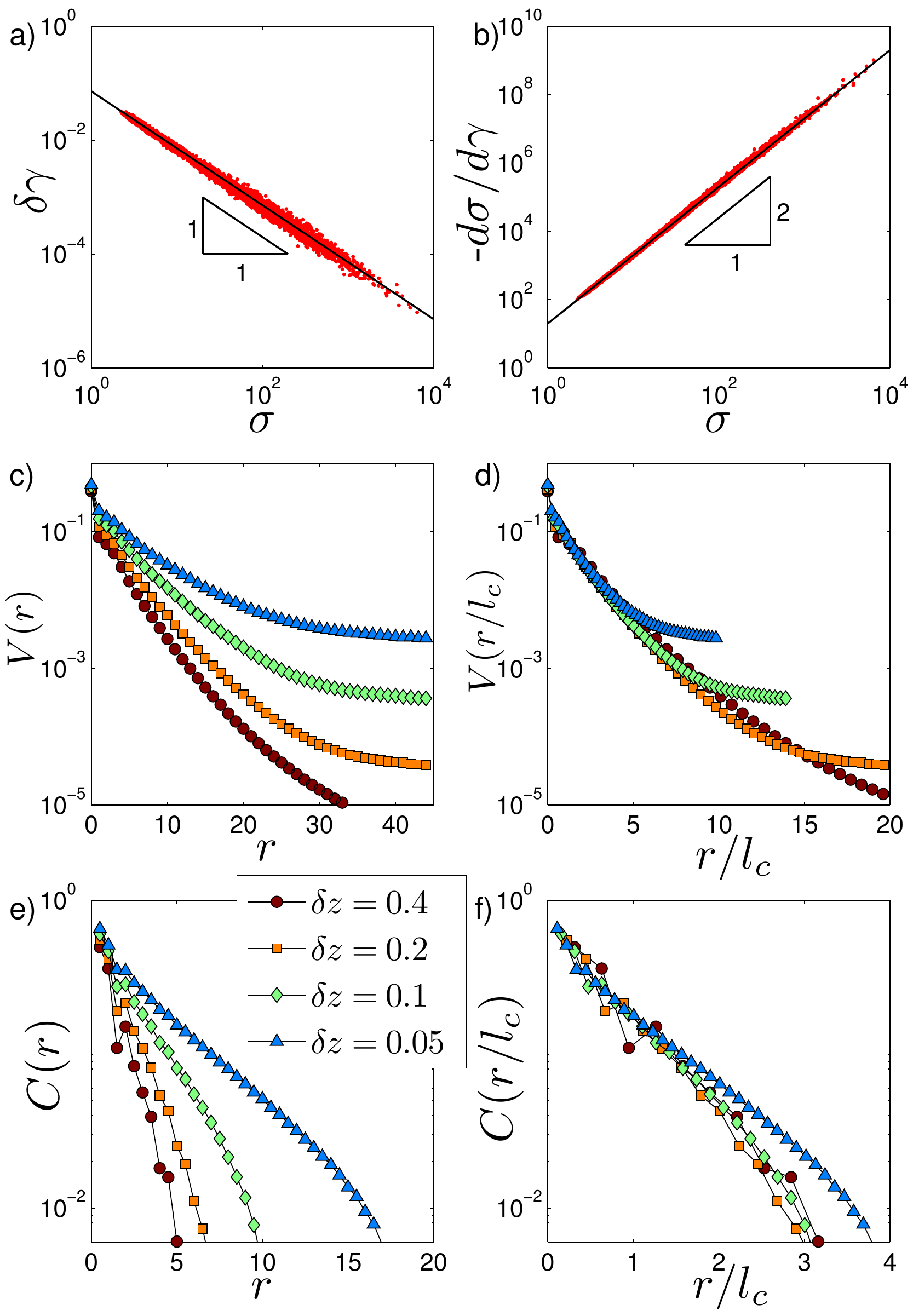}}
\caption{ (Color online) Hard-particle suspensions modeled by the ASM simulated as in \cite{lernerpnas}. {\bf a)}~Distance in strain $\delta \gamma$ to the SIJ point (as explained in the text) vs.~the stress $\sigma$. {\bf b)}~Scaling of  $-\frac{d\sigma}{d \gamma}$ vs $\sigma$ for hard-particle systems under shear flow. {\bf c)} Average velocity response to a local perturbation $V(r)$ vs.~the distance $r$ to the perturbation. {\bf d)} Rescaled average response  $V(r)/\delta z$ vs.~the rescaled distance $r/l_c$. {\bf e)} Velocity correlation $C(r) \equiv \langle\vec{V}_i\cdot\vec{V}_j\rangle/\langle V^2 \rangle$ vs the distance $r$ between particles $i$ and $j$. {\bf f)} Velocity correlations $C(r)$ vs the rescaled distance $r/l_c$.\label{flowProp}   }
\end{center}
\end{figure}

\section{Conclusion}
We have shown that the strain-induced jamming transition  is akin to a critical point, but of a curious kind: when this transition is approached,  exponents  depend on the initial conditions, and two length scales diverge. 
When going away from this transition, the associated buckling behavior becomes universal and only one singular length scale plays a role.  Some of these results could help  distinguishing which processes control the non-linear behavior of floppy gels.  On the one hand, although our models are much simpler than gels of biopolymers, we expect our prediction that the non-affine displacements of the network nodes spike when stiffening occurs to be robust; in fact, this was observed numerically in models that include bending elasticity 
 \cite{giessen} and has been recently  related to the SIJ transition \cite{Mackintosh_PRE_2012}.  This prediction could be used to test when this scenario is responsible for  the observed shear-stiffening of these materials, or when non-linearities of the individual fibers are instead responsible, as initially proposed \cite{gel}.  On the other hand, the presence of weak interactions (e.g. bending) was shown to affect qualitatively some aspects of the elastic response in floppy materials \cite{during12}, and should be included in the future to obtain a description of the SIJ transition that applies quantitatively to gels. 

Most importantly, we have argued that this same transition plays a role in simple models of suspensions flows (where weak interactions are absent), thus building a bridge between two quite distinct classes of materials. This approach has allowed us to characterize quantitatively the ``buckling of force chains" and to propose a new perspective on the self-organization characterizing  these driven materials. Our approach predicts a rather subtle scenario with a vanishing strain scale and two diverging length scales, whose dominant one diverges weakly as $l_c\sim 1/\sqrt\delta z\approx p^{-0.18}$, and is decoupled from the divergence of the viscosity.  Our numerical observations strongly support this prediction. Our exponent for the divergence of $l_c$ is much smaller than early numerical estimates \cite{olsson,heussinger2009}. However, these numerics were performed in relatively small systems, and the length was measured as a function of $\phi$ instead of $p$, which considerably increases finite size effects. Although the presence of hydrodynamic interactions complicates the extraction of length scales from velocity correlations  \cite{martin2010}, measurements of the response to local perturbations, as well as studies of the non-local rheology \cite{Henann,bouzid} should enable to test our predictions experimentally.

\begin{acknowledgments}
We thank Eric DeGiuli for constructive comments
on the manuscript. This work has been supported by the Sloan Fellowship,
National Science Foundation DMR-1105387, Petroleum Research Fund
52031-DNI9 and by the MRSEC Program of the National Science Foundation
DMR-0820341. G.D. acknowledges partial support from  CONICYT  PAI/Apoyo al Retorno 82130057. 
\end{acknowledgments}






\appendix
 \section{perturbation expansion}
In order to find an expression for the evolution we need to solve the force balance Eq.(\ref{forcebalance1}), which can be written as
\begin{equation}
\mathcal{N} \ket{\tau}=-\ket{v_{\rm f}},
\label{inverse}
\end{equation}
where $\langle \alpha \ket{v_{\rm f}}=f_{\rm ext}\equiv \cos(2\theta_\alpha)$ and the $\mathcal{N}$-matrix is defined in the main text Eq.(\ref{N1D}).  In the linear chain the critical point corresponds to a line (i.e.~$\theta_\alpha=0$ for all $\alpha$), signaling that the natural small parameter should be small angles. A standard perturbation analysis around the critical point cannot be carried out because it is singular. Therefore, we introduce the parameter $\kappa$ and write the $\mathcal{N}$-matrix as $\mathcal{N}=\mathcal{N}_\kappa+\delta \mathcal{N}_\kappa$ where
\begin{eqnarray}
\bra{\alpha}\mathcal{N}_\kappa\ket{\beta}=2\delta_{\alpha, \beta}-(1-\kappa)\delta_{\beta, \alpha+1} -(1-\kappa)\delta_{\beta, \alpha-1} \\
\bra{\alpha}\delta\mathcal{N}_\kappa\ket{\beta}=(\frac{1}{2}(\theta_\alpha-\theta_\beta)^2-\kappa)(\delta_{\beta, \alpha+1} +\delta_{\beta, \alpha-1} ),
\end{eqnarray}
and $\kappa$ will be chosen to remove the singularity.
Equation (\ref{inverse})  then reads as $ \mathcal{N}_\kappa \ket{\tau}=-\ket{v_{\rm f}}-\delta\mathcal{N}_\kappa \ket{\tau}$ for which a solution can be formally written as a series
\begin{multline}
 \ket{\tau}=-\mathcal{N}_\kappa^{-1}\ket{v_{\rm f}}+\mathcal{N}_\kappa^{-1}\delta\mathcal{N}_\kappa \mathcal{N}_\kappa^{-1}\ket{v_{\rm f}}\\-\mathcal{N}_\kappa^{-1}\delta\mathcal{N}_\kappa \mathcal{N}_\kappa^{-1}\delta\mathcal{N}_\kappa\mathcal{N}_\kappa^{-1}\ket{v_{\rm f}}+\ldots
\label{expansion}
\end{multline}
To calculate the contribution of the higher order terms in the expansion (\ref{expansion}) we use the eigenvectors and eigenvalues of $\mathcal{N}_\kappa$, which  are 
given by $$ \langle\alpha\ket{\psi_q}=\frac{e^{i2\pi q \alpha/N}}{\sqrt{N}}\,\,\,\, {\rm and}\,\,\,\, \omega^2_q=2(1-(1-\kappa)\cos(2\pi q/N)),$$ respectively. $q$ is an integer between $[0,N-1]$ and the contraction 
\begin{equation}
\bra{\psi_q} \delta\mathcal{N}_\kappa \ket{\psi_l}=(e^{2\pi il/N}+e^{-2\pi iq/N})\left(\frac{1}{2N}\Delta_{l-q}-\kappa\delta_{q,l}\right)\nonumber
\end{equation}
where $\Delta_q=\sum^{N-1}_{\alpha=0} (\theta_\alpha-\theta_{\alpha+1})^2e^{2\pi iq\alpha/N}$ is the Fourier transform of $(\theta_\alpha-\theta_{\alpha+1})^2$.   In the thermodynamic limit $\vert\Delta_q\vert^2/N$ is singular at $q=0$ due to the non zero mean of the real space function. This singularity is precisely what we need to remove by taking $$\kappa=\Delta_{q=0}=\frac{1}{2N}\sum^{N-1}_{\alpha=0} (\theta_\alpha-\theta_{\alpha+1})^2=\langle \Delta\theta^2\rangle/2.$$
The minimal frequency $\omega^2_{0}=2\kappa$ is given by the mean square of the angle differences between neighboring nodes. Then, $\mathcal{N}_\kappa$ corresponds precisely to the average  $\langle\mathcal{N}\rangle$, where the spatial average is taken over the coefficients of the  $\mathcal{N}$-matrix.

Close to the critical point $\theta_\alpha$ can be considered small and $\ket{v_{\rm f}}\approx\sqrt{N}\ket{\psi_0}$, therefore the first term in the series (\ref{expansion}) leads to a constant tension $\tau_\alpha= -1/\langle \Delta\theta^2\rangle$ which corresponds to the average tension $\langle \tau\rangle$.  In order to get an asymptotic expansion, the square of the average tension $ \langle\tau\rangle^2 \equiv\bra{v_{\rm f}}\mathcal{N}_\kappa^{-2}\ket{v_{\rm f}}$ must be larger than the amplitude of the fluctuations. From the expansion (\ref{expansion}) the amplitude of the first correction to the average is given by
\begin{eqnarray}
\delta \tau^2&=&\bra{v_{\rm f}}\mathcal{N}_\kappa^{-1}\delta\mathcal{N}_\kappa\mathcal{N}_\kappa^{-2}\delta\mathcal{N}_\kappa \mathcal{N}_\kappa^{-1}\ket{v_{\rm f}}.
\end{eqnarray}
The approximation is only valid if the ratio of the fluctuations amplitude over the average amplitude satisfies
\begin{equation}
\frac{\delta\tau^2}{\langle\tau\rangle^2}=\sum_{q\neq 0}\frac{\vert\bra{\psi_q}\delta\mathcal{N}_\kappa\ket{\psi_0}\vert^2}{\omega^4_q}\ll1.
\end{equation}
A straightforward calculation leads to the expression for the overlap 
\begin{multline}
\vert\bra{\psi_q}\delta\mathcal{N}_\kappa\ket{\psi_0}\vert^2=\frac{16}{N^3}\cos(\pi \frac{q}{N})^2\sum_{k_1 k_2}\theta_{k_2-q}\theta_{-k_2}\theta_{k_1+q}\theta_{k_1}\\
\times\sin(\pi\frac{k_2-q}{N})\sin(\pi \frac{k_2}{N})\sin(\pi\frac{k_1+q}{N})\sin(\pi \frac{k_1}{N}).\nonumber
\end{multline}
The final result depends on the distribution of the initial conditions for the angles $\theta_k(0)$. Considering distributions with small or no higher order cumulant, the distribution can be taken as a Gaussian. In this case the thermodynamic limit of the ratio $\frac{\delta\tau^2}{\langle\tau\rangle^2}$ is given by
\begin{multline}
\frac{\delta\tau^2}{\langle\tau\rangle^2}\approx32c_\mu^2\int_0^1\frac{dkdq}{\omega_q^4}\cos(\pi q)^2\left[\sin(\pi(k-q))\sin(\pi k)\right]^{\mu+2}\\ 
\times e^{-2(\nu_k+\nu_{k-q})\int_0^\gamma\frac{d\tau}{\langle \Delta\theta^2\rangle}},
\label{ratio}
\end{multline}
where we  used  $q/N\rightarrow q$ with $q\in[0,1]$. Eq.(\ref{ratio}) results from the diffusion equation (\ref{diffusionEquation}) and the statistical space homogeneity of the system. Notice that for the case of an initial distribution given by random independent angles $\theta_\alpha(0)$ the last expression is exact since it is Gaussian distributed. Finally, an upper bound can be obtained for Eq.(\ref{ratio}), taking $q=0$ except at the denominator $\omega^4_q$, which reads $$ \frac{\delta\tau^2}{\langle\tau\rangle^2} \lesssim \langle \Delta\theta^2\rangle^{\frac{1+\mu}{6+2\mu}}\sim (\gamma_c-\gamma)^{1/2}.$$  
 Thus close to the critical point the approximation is valid and fluctuations can be neglected to a first approximation. An asymptotic behavior is expected for the higher order contributions in the expansion (\ref{expansion}), but will be considered in detail elsewhere.

\section{The effect of noise and the loss of reversibility }
We shall consider the diffusion equation with a $\delta$-correlated gaussian noise 
\begin{equation}
\dot\theta_\alpha=\frac{\dot\gamma}{\langle \Delta\theta^2\rangle}(\theta_{\alpha+1}-2\theta_\alpha+\theta_{\alpha-1})+\xi_\alpha(t)
\label{diffusionEquationNoise}
\end{equation} 
where $\langle \xi_\alpha(t)\xi_\beta(t')\rangle=\varsigma^2 \delta_{\alpha,\beta} \delta(t-t')$.   Notice that in this Appendix for sake of simplicity we will not use a dimensionless time. Taking the Fourier transform we obtain 
\begin{eqnarray}
\theta_k(t)=\theta_k(t_0)\exp[ \dot\gamma{\nu_k}\int_{t_0}^t \frac{d\tau}{\langle \Delta\theta^2\rangle}]\nonumber\\
+\int_{t_0}^t \exp[ \dot\gamma{\nu_k}\int_{u}^t \frac{d\tau}{\langle \Delta\theta^2\rangle}] \xi_k d{u}
\label{evol}
\end{eqnarray}
with the Fourier components $\xi_k=\sum^{N-1}_{\alpha=0}\xi_\alpha e^{-i2\pi k\alpha/N}$.  The self consistent equation (\ref{self}) is now given by
\begin{eqnarray}
\langle \Delta\theta^2\rangle=\frac{4}{N}\sum_{k} g(k)e^{2\dot\gamma\nu_k\int_{t_0}^t \frac{d\tau}{\langle \Delta\theta^2\rangle}}\sin[\pi  k/N]^2\nonumber\\
+\varsigma^2\frac{4}{N}\sum_{k} \int_{t_0}^t e^{ 2\dot\gamma{\nu_k}\int_{u}^t \frac{d\tau}{\langle \Delta\theta^2\rangle}}  \sin[\pi  k/N]^2du.
\label{self2}
\end{eqnarray}
In the last expression we average over the noise $\xi$ and over the initial conditions $\langle\vert \theta_k(t_0)\vert^2\rangle/N=g(k)$. In deriving the last equation we have assumed that $\langle \Delta\theta^2\rangle$ is self-averaging which implies that it is not correlated with $\xi_k$ or $\theta_k(t_0)$ in the thermodynamic limit $N\rightarrow\infty$.In this limit we can replace $\frac{1}{N}\sum \rightarrow \int$ and $k/N\rightarrow k$. Considering  $g(k)=\epsilon^2 c_\mu\vert\sin(\pi k)\vert^{\mu}$ Eq.(\ref{self}) reads 
\begin{eqnarray}
\langle \Delta\theta^2\rangle&=&\epsilon^24c_\mu\int^1_0e^{2\dot\gamma\nu_k\int_{t_0}^t \frac{d\tau}{\langle \Delta\theta^2\rangle}}\sin[\pi  k]^{\mu+2}dk\nonumber\\
&&+\varsigma^2 4\int_{t_0}^t\int_0^1  e^{ 2\dot\gamma{\nu_k}\int_{u}^t\frac{d\tau}{\langle \Delta\theta^2\rangle}}  \sin[\pi  k]^2dkdu.
\label{self3}
\end{eqnarray}
A  bound for the contribution of the noise can be calculated by extending the lower  limit of integration in the exponent from $u\rightarrow t_0$. This bound shows that the term induced by the noise can be neglected in the evolution for times $(t-t_0)\ll c_\mu \epsilon^2/\varsigma^2$, which is equivalent to the condition that $\Delta \gamma/\dot\gamma \ll c_\mu \epsilon^2/\varsigma^2$. If the noise is considered to be introduced by thermal fluctuations, the required condition is that the system is strained fast enough toward the critical point  to prevent  thermalization.  

A very different situation occurs under the reverse evolution from the critical point, i.e.~the buckling process.  We shall consider that  the system is evolved until a time $t_f$ smaller than the critical time, and then the dynamics is reversed, which means changing $\dot{\gamma}\rightarrow-\dot{\gamma}$ in the evolution equation (\ref{evol}). To simplify the backward evolution we will consider that there is no noise during the evolution, hence $\varsigma=0$ and the only source of irreversibility comes from the accumulated noise in the forward evolution. The amplitude of the accumulated noise $\varepsilon^2$ must be much smaller that $\langle \Delta\theta^2\rangle$ to ensure that the system is non-Brownian.  Depending on the type of noise present in the evolution towards the critical point the final deviation from the exact result might depend on the wave-number. However, it can be shown that in most cases the particular choice of noise makes no difference for the backward evolution, so we consider for simplicity that the accumulated  error is of the same order for every mode. Therfore, the initial condition is given by $\theta_k(t_f)=\theta_k(t_0)\exp[ \dot\gamma{\nu_k}\int_{t_0}^{t_f} \frac{d\tau}{\langle \Delta\theta^2\rangle}]+\varepsilon$, and the reverse self-consistent equation is finally given by 
\begin{eqnarray}
\langle \Delta\theta^2\rangle&=&4\int^1_0g(k) e^{2\dot\gamma\nu_k\left(\int_{0}^{t_f}\frac{d\tau}{\langle \Delta\theta^2\rangle}-\int_{t_f}^t \frac{d\tau}{\langle \Delta\theta^2\rangle}\right)}\sin[\pi  k]^{2}dk\nonumber\\
&&+\varepsilon^2  4\int_0^1  e^{ -2\dot\gamma{\nu_k}\int_{t_f}^t \frac{d\tau}{\langle \Delta\theta^2\rangle}}  \sin[\pi  k]^2dk.
\label{self5}
\end{eqnarray}
Initially the backward evolution is dominated by the reversible term, assuming that the noise amplitude $\varepsilon^2\ll \langle \Delta\theta^2\rangle$. Then the first term on the RHS of Eq.(\ref{self5}) is proportional to $ \epsilon^{\frac{-4}{1+\mu}} a_\mu (\gamma_c-\gamma)^{\frac{3+\mu}{1+\mu}}$ where $\gamma=\dot\gamma(2t_f-t)$  for $t>t_f$ and $a_\mu$ is a constant that only depends on $\mu$ ($a_\mu$ can be expressed in terms of Gamma functions).
We now estimate the time $t_{\rm noise}$ at which the noise term becomes of the same order of the reversible term:
\begin{equation}
\epsilon^{\frac{-4}{1+\mu}} a_\mu(\gamma_c-\gamma)^{\frac{3+\mu}{1+\mu}}\sim4\varepsilon^2  \int_0^1  e^{ -2{\nu_k}\rho(t)}  \sin[\pi  k]^2dk, 
\label{cross}
\end{equation} 
where $\rho(t)=\dot{\gamma}\int_{t_f}^t \frac{d\tau}{\langle \Delta\theta^2\rangle}$. In this regime, while $t\ll t_{\rm noise}$ we can calculate $\rho(t)$ using the reversible evolution of $\langle \Delta\theta^2\rangle$. In terms of the dimensionless strain $\gamma$ one gets 
\begin{equation}
\rho(t)\equiv\rho_{\rm r}(\gamma)=  \frac{\epsilon^{\frac{4}{1+\mu}}(\mu+1)}{2a_\mu(\gamma_c-\gamma_f)^{\frac{2}{1+\mu}}}\left[1-\left(\frac{\gamma_c-\gamma}{\gamma_c-\gamma_f}\right)^{-\frac{2}{1+\mu}}\right],
\label{rho}
\end{equation}
where $\gamma_f=\dot\gamma t_f$. The integral in Eq.(\ref{cross}) can be expressed as:
\begin{equation}
\int_0^1  e^{ -2{\nu_k}\rho_{\rm r}(\gamma)}  \sin[\pi  k]^2dk\equiv\frac{e^{4\rho_{\rm r}(\gamma)}}{2}\left[I_0(4\rho_{\rm r}(\gamma))+I_1(4\rho_{\rm r}(\gamma))\right],
\label{Bessel}
\end{equation}
where $I_0(x)$  and  $I_1(x)$ are modified Bessel functions. Since we are considering exponentially small noise we need to reach values of $\rho(\gamma)\gg1$.  In this limit we can  use the asymptotic expression for the Bessel function which leads to the equation:
\begin{equation}
\epsilon^{\frac{-4}{1+\mu}} a_\mu(\gamma_c-\gamma)^{\frac{3+\mu}{1+\mu}}\sim4\varepsilon^2  e^{8 \rho_{\rm r}(\gamma)}/\sqrt{8\pi \rho_{\rm r}(\gamma)}.
\label{cross}
\end{equation} 
The crossover strain $\gamma_{\rm noise}$ must take place when $  \rho_{\rm r}(\gamma_{\rm noise})\sim \vert\ln(\varepsilon)\vert/8$ which leads to the result $(\gamma_c-\gamma_{\rm noise})\sim (\rho_{\rm r}(-\infty)-\vert\log(\varepsilon)\vert/4)^{-\frac{1+\mu}{2}}$.
The reversibility is then only lost if $\frac{\vert\log(\varepsilon)\vert}{4\rho_{\rm r}(-\infty)}<1$, however, due to the logarithmic dependence in the noise this inequality is easily satisfied close to the critical point. If $\frac{\vert\log(\varepsilon)\vert}{4\rho_{\rm r}(-\infty)}\ll1$ one finds that the crossover time $t_{\rm noise}\sim t_f +\langle \Delta\theta(\gamma_f)^2\rangle \vert\log(\varepsilon)\vert/\dot\gamma$ above which the noise term becomes dominant and  the reversible term can be neglected. Then the integral equation (\ref{self5}) can be written as:
\begin{eqnarray}
\langle \Delta\theta^2\rangle&=&\varepsilon^2 4\int_0^1  e^{ -2\dot\gamma{\nu_k}\int_{t_f}^t \frac{d\tau}{\langle \Delta\theta^2\rangle}}  \sin[\pi  k]^2dk\nonumber\\
&=&\varepsilon^2  2 e^{4\rho(t)}(I_0(4\rho(t))+I_1(4\rho(t))).
\label{selfAs}
\end{eqnarray}
The integral form of $\rho(t)$ allow us to write  $$\rho(t)=\rho_{\rm r}(\gamma_{\rm noise})+\dot\gamma\int_{t_{\rm noise}}^t \frac{d\tau}{\langle \Delta\theta^2\rangle}.$$ Since $\rho_{\rm r}(\gamma_{\rm noise})\gg1$  we can use the asymptotic behavior for the modified Bessel functions in Eq.(\ref{selfAs}) which leads to  
\begin{multline}
\langle \Delta\theta^2\rangle\approx4\varepsilon^2  \frac{e^{8\rho_{\rm r}(\gamma_{\rm noise})+8\dot\gamma\int_{t_{\rm noise}}^t \frac{d\tau}{\langle \Delta\theta^2\rangle}}}{\sqrt{8\pi  \rho_{\rm r}(\gamma_{\rm noise})}}\\=\langle \Delta\theta(t_{\rm noise})^2\rangle e^{8\dot\gamma\int_{t_{\rm noise}}^t \frac{d\tau}{\langle \Delta\theta^2\rangle}}.
\end{multline}
Finally we obtain the asymptotic solution for the reverse evolution $$\langle \Delta\theta^2\rangle\sim (\tilde\gamma_c-\gamma),$$  where the effective critical strain depends on the initial conditions $\tilde\gamma_c\approx \gamma_{\rm noise}+\langle \Delta\theta(t_{\rm noise})^2\rangle/8$, but the scaling exponent is universal.

\section{Scaling of $\omega_{\rm min}$ and $\sigma$ with strain}
We consider the ASM framework \cite{lernerpnas} in which overdamped hard spheres are driven under shear flow, and adopt the notations for dimensionless forces and velocities from the main text. The linear operator ${\cal S}$ and its transpose ${\cal S}^T$ are defined via the following relations
\begin{eqnarray}
\vec{n}_{ij}\cdot(\vec{V}_j - \vec{V}_i) &\longleftrightarrow & {\cal S}\ket{V}\ , \\
\sum_{i(j)}\vec{n}_{ij}\tau_{ij} &\longleftrightarrow & {\cal S}^T\ket{\tau}\ , 
\end{eqnarray}
where $\vec{n}_{ij}$ are the unit vectors pointing from the center of particle $i$ to the center of particle $j$, and 
$i(j)$ are all the particles $i$ that are in contact with particle $j$. We consider the limit of perfectly rigid particles, therefore the velocities $\ket{V}$ must preserve the distance between the centers of every pair of particles that are in contact, i.e.~${\cal S}\ket{V}=0$. Decomposing the velocities to the driving affine flow and the nonaffine part $\ket{V} = \ket{V^{\rm f}} + \ket{V^{\rm na}}$, we require that drag forces $\ket{F^{\rm drag}} = \ket{V^{\rm f}} - \ket{V}$ balance the net forces on particles due to contacts 
\begin{equation}\label{foo03}
{\cal S}^T\ket{\tau} = -\ket{F^{\rm drag}}  = \ket{V} - \ket{V^{\rm f}}.
\end{equation}
Operating on the above equation with ${\cal S}$ leads to Eq.(\ref{forcehigh1}). Inverting for the contact forces, we obtain 
\begin{equation}\label{foo02}
\ket{\tau} = -{\cal N}^{-1}{\cal S}\ket{V^{\rm f}},
\end{equation}
where ${\cal N} \equiv {\cal S}{\cal S}^T$. Inserting Eq.(\ref{foo02}) into Eq.(\ref{foo03}) results in the non-affine velocities
\begin{equation}
\ket{V^{\rm na}} =  -{\cal S}^T{\cal N}^{-1}{\cal S}\ket{V^{\rm f}}.
\end{equation}
Denoting the $k^{\rm th}$ particles' coordinates by $\vec{R}_k$, and the differences $\vec{R}_{ij} \equiv \vec{R}_j - \vec{R}_i$, the variation of the pairwise distance $r_\alpha \equiv \sqrt{\vec{R}_\alpha\cdot\vec{R}_\alpha}$ for the contact $\alpha = \langle ij \rangle$ under simple shear in the $x-y$ plane is given by 
\begin{equation}
\bra{\alpha}{\cal S}\ket{V^{\rm f}} = \frac{\partial r_\alpha}{\partial \gamma} = \frac{ \hat{x}\cdot\vec{R}_\alpha\vec{R}_\alpha\cdot\hat{y}}{r_\alpha}\ ,
\end{equation}
where $\hat{x},\hat{y}$ denote Cartesian unit vectors. From here the stress is given by
\begin{equation}\label{foo01}
\sigma = -\frac{\bra{\tau}{\cal S}\ket{V^{\rm f}}}{\Omega} = 
\frac{\bra{V^{\rm f}}{\cal S}^T{\cal N}^{-1}{\cal S}\ket{V^{\rm f}}}{\Omega} = 
\frac{\bra{V^{\rm na}}V^{\rm na}\rangle}{\Omega},
\end{equation}
where $\Omega$ is the dimensionless volume.

In our framework, total derivatives of a regular observable~$O$ are taken as $\frac{dO}{d\gamma} = \frac{\partial O}{\partial \gamma} + \bra{\frac{\partial O}{\partial R}}V^{\rm na}\rangle$. As $\frac{\partial O}{\partial \gamma}$ is always regular, whereas close to the critical point $\ket{V^{\rm na}}$ is singular \cite{lernerpnas}, we approximate in the following $\frac{dO}{d\gamma} \simeq \bra{\frac{\partial O}{\partial R}}V^{\rm na}\rangle$. We let $\ket{W}$ be a vector in the space of particles and calculate
\begin{equation}
\frac{d{\cal S}}{d\gamma}\ket{W} \simeq \ket{(\bra{V^{\rm na}}{\cal K}\ket{W})}\ ,
\end{equation}
where the contraction of ${\cal K}$ on a pair of any vectors $\ket{U}$,$\ket{W}$ is a vector \emph{in the space of contacts} with components defined~via
\begin{equation}
\vec{U}_{ij}\cdot\left( \frac{\stackrel{\leftrightarrow}{\cal I}}{r_{ij}} - \frac{\vec{n}_{ij}\vec{n}_{ij}}{r_{ij}}\right)\cdot
\vec{W}_{ij} \quad \longleftrightarrow \quad \bra{U}{\cal K}\ket{W},
\end{equation}
where $\vec{U}_{ij}\equiv \vec{U}_j - \vec{U}_i$, and $\stackrel{\leftrightarrow}{\cal I}$ denotes the unit tensor. Notice that components of the vector $\ket{ \left( \bra{W}{\cal K}\ket{W}\right) }$ are non-negative for any vector $\ket{W}$. 
Denoting by $\ket{\phi_{\rm min}}$ the normalized eigenvector of ${\cal N}$ associated to the minimal eigenvalue $\omega_{\rm min}^2 = \bra{\phi_{\rm min}}{\cal N}\ket{\phi_{\rm min}}$, the derivative of $\omega_{\rm min}^2$ with respect to strain is calculated as
\begin{equation}\label{foo05}
\frac{d(\omega_{\rm min}^2)}{d\gamma} \simeq 2\bra{\phi_{\rm min}}(\bra{V^{\rm na}}{\cal K}{\cal S}^T\ket{\phi_{\rm min}})\rangle,
\end{equation}
since $\bra{\phi_{\rm min}}\frac{d\phi_{\rm min}}{d\gamma}\rangle = 0$ from normalization. Close to the critical point the minimal mode has a finite coupling to the shear \cite{Lerner2012} (this simply comes from the fact that even for frictionless particles, at jamming the stress has a finite anisotropy, i.e. $\mu=\sigma/p>0$), i.e.~$\bra{\phi_{\rm min}}{\cal S}\ket{V^{\rm f}} \sim||\phi_{\rm min}||\times||{\cal S}\ket{V^{\rm f}} || = ||{\cal S}\ket{V^{\rm f}} || \sim \sqrt{N}$ (the last relation simply stem from the fact that the components of ${\cal S}\ket{V^{\rm f}}$ are of order one in our dimensionless units). Contact forces $\ket{\tau}$ (see Eq.(\ref{foo02})) are thus dominated by the minimal mode, i.e.
\begin{equation}\label{foo06}
\ket{\tau} \sim -\frac{\bra{\phi_{\rm min}}{\cal S}\ket{V^{\rm f}}\ket{\phi_{\rm min}}}{\omega_{\rm min}^2}\sim\frac{\sqrt{N}}{\omega_{\rm min}^2}\ket{\phi_{\rm min}}.
\end{equation}
Operating on the above relation with ${\cal S}$ and using Eq.(\ref{foo03}), we find
\begin{equation}
{\cal S}^T\ket{\phi_{\rm min}} \sim \frac{\omega_{\rm min}^2}{\sqrt{N}}{\cal S}^T\ket{\tau} \sim \frac{\omega_{\rm min}^2}{\sqrt{N}}\ket{V^{\rm na}}.
\end{equation}
Inserting this expression for ${\cal S}^T\ket{\phi_{\rm min}}$ in Eq.(\ref{foo05}), we obtain
\begin{equation}\label{foo08}
\frac{d(\omega_{\rm min}^2)}{d\gamma} \sim \omega_{\rm min}^2 
 \frac{\bra{\phi_{\rm min}}\left(\bra{V^{\rm na}}{\cal K}\ket{V^{\rm na}}\right)\rangle}{\sqrt{N}}.
\end{equation}
Since contact forces are positive for repulsive particles, close to the critical point where contact forces are dominated by the minimal mode, all components of $\ket{\phi_{\rm min}}$ are non-negative. This is also true for the components of $\ket{\left( \bra{V^{\rm na}}{\cal K}\ket{V^{\rm na}}\right)}$ by construction. Thus we expect for the overlap $\bra{\phi_{\rm min}}\left(\bra{V^{\rm na}}{\cal K}\ket{V^{\rm na}}\right)\rangle\sim ||\phi_{\rm min}||\times||\bra{V^{\rm na}}{\cal K}\ket{V^{\rm na}}||=||\bra{V^{\rm na}}{\cal K}\ket{V^{\rm na}}||$. We finally assume that  the {\it relative} non-affine displacement, of order $(||\bra{V^{\rm na}}{\cal K}\ket{V^{\rm na}}||/\sqrt{N})^{1/2}$, are the order of  the relative non-affine velocities between particles, by definition of order  $||V^{\rm na}||/\sqrt{N}$. This assumption is consistent with our results on the buckling of the one-dimensional chain, where it applies. This assumption implies that 
\begin{equation}\label{foo07}
\frac{\bra{\phi_{\rm min}}\left(\bra{V^{\rm na}}{\cal K}\ket{V^{\rm na}}\right)\rangle}{\sqrt{N}} \sim \frac{||V^{\rm na}||^2}{N}.
\end{equation}
Returning to Eq.(\ref{foo01}), we use Eq.(\ref{foo06}) for the forces to obtain
\begin{equation}\label{foo04}
\sigma \sim \frac{||V^{\rm na}||^2}{\Omega} \sim \frac{\bra{\phi_{\rm min}}{\cal S}\ket{V^{\rm f}}^2}{\Omega \omega_{\rm min}^2}\sim\frac{N}{\Omega\omega_{\rm min}^2}\sim\frac{1}{\omega_{\rm min}^2}.
\end{equation}
Finally, using Eqs.(\ref{foo07},\ref{foo04}) in Eq.(\ref{foo08}), we find that $\frac{d(\omega_{\rm min}^2)}{d\gamma} \sim {\cal O}(1)$ or $\omega_{\rm min} \sim \sqrt{\gamma - \gamma_c}$ with $\gamma_c < \gamma$. 

A straightforward consequence of the scalings $\omega_{\rm min} \sim \sqrt{\gamma - \gamma_c}$ and 
$\sigma \sim 1/\omega_{\rm min}^2$ (see Eq.(\ref{foo04})) is a prediction for the relaxation of the stress $\sigma \sim 1/(\gamma - \gamma_c)$, or alternatively
\begin{equation}
\frac{d\sigma}{d\gamma} \sim -\sigma^2\ ,
\end{equation}
which is validated numerically in Fig.(\ref{flowProp},b).

\begin{figure}[t]
\begin{center}
\centerline{\includegraphics[width=0.5\textwidth]{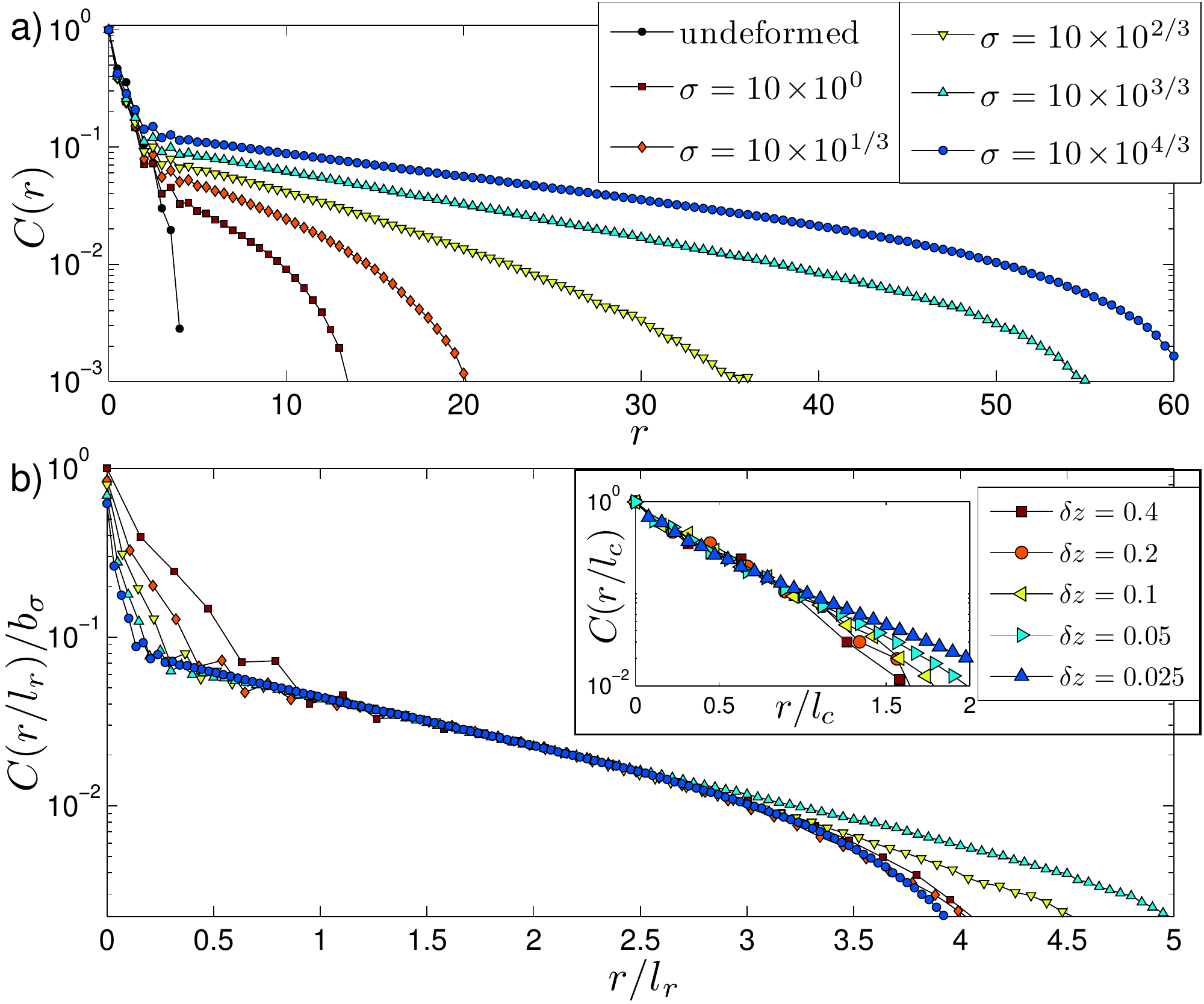}}
\caption{ (Color online) Velocity correlations $C(r)$ for networks under shear. Panel a) shows that the raw correlation functions have an initial exponential decay on the lengthscale $l_c \sim 1/\sqrt{\delta z}$. We also show the velocity correlations measured in an undeformed, isotropic network of the same coordination (black dots), demonstrating that the initial decay occurs on the same lengthscale, independent of the stress. Panel b): the correlation functions rescaled by constants $b_\sigma$ which are chosen such that the functions collapse with the second lengthscale $l_r$, scaling as $l_r\sim\sqrt{\sigma}$ at fixed $z$. We find numerically $b_\sigma \sim \log(\sigma)$. In the inset of panel b) we plot the velocity correlations measured in undeformed, isotropic networks with various coordinations. \label{VeloCorr}}
\end{center}
\end{figure}

\section{Numerical methods}
Data for the flow of hard particles within the ASM framework \cite{lernerpnas} is generated using the algorithm described in \cite{simulation_paper}. We have simulated a binary mixture of large and small discs in two-dimensions with $N=4096$ particles, setting the ratio of radii of large and small particles to 1.4. Systems were deformed under simple shear with Lees-Edwards periodic boundary conditions for strains of at least 300\% before statistics were collected. We simulated systems at various packing fractions ranging from $\phi=0.820$ to $\phi=0.840$.

The same algorithm described in \cite{simulation_paper} is straightforwardly adapted for deforming networks of rigid rods by keeping the topology of the networks fixed. Initial floppy networks of rods with coordination $z<2d$ were created as described in \cite{Wyartmaha}. We have simulated networks of $N=4096$ and $N=40000$ nodes, and coordinations varying between $z=3.6$ and $z=3.95$. 
Networks were deformed under simple shear with Lees-Edwards periodic boundary conditions until the dimensionless stress reached $\sigma \approx 10^6$. Then, the shear was reversed (see for example the stress-strain signal of Fig.(\ref{reverseStrain},a)), and data was collected. Responses to point perturbations and velocity correlations were measured in the larger networks, whereas the spectral analysis as displayed in Fig.(\ref{Density}) was performed on the smaller networks. The response to a point perturbation (Figs.(\ref{linePro},c),(\ref{VelResponse}),(\ref{flowProp},c,d)) at a bond $\alpha$ is calculated by solving the equation ${\cal N}\ket{f} = \ket{\alpha}$ for $\ket{f}$, and then calculating $\ket{V} = {\cal S}^T\ket{f}$ (see Appendix~C for definitions of ${\cal S},{\cal N}$). The derivatives $d\sigma/d\gamma$ were calculated by finite differences.

Linear chains with $N=10000$ nodes we deformed under extensional flow with periodic boundary conditions using the same method used for random floppy networks.

\section{Velocity correlations in strained random networks}

In Fig.(\ref{VeloCorr}) the velocity correlations $C(r) \equiv \frac{\langle\vec{V}_i\cdot\vec{V}_j\rangle}{\langle V^2\rangle}$ \emph{vs.} distance $r$ between particle $i$ and $j$ is plotted for networks of $N=40,000$ nodes with $\delta z = 0.2$ in two dimensions, at various stresses. We find that $C(r)$ initially decays exponentially on the scale $l_c\sim1/\sqrt{\delta z}$, and then crosses over to an exponential decay on the lengthscale $l_r \sim \sqrt{\sigma}$. These correlation functions are similar to the velocity response to a point perturbation presented in Fig.(\ref{VelResponse}). Notice in particular the two exponential decays, on the lengthscale $l_c$ at short distance  and $l_r$ in the far field.

\bibliography{reference8}{}

\end{document}